\documentstyle[prb,twocolumn,aps,epsf,psfrag]{revtex}

\begin{document}
\draft

\twocolumn[
\hsize\textwidth\columnwidth\hsize\csname
@twocolumnfalse\endcsname

\title{Spin dynamics of Mn$_{12}$-acetate in the thermally-activated \\
  tunneling regime: ac-susceptibility and magnetization relaxation}

\author{Teemu Pohjola$^{1,2}$ and Herbert Schoeller$^3$}

\address{$^1$Materials Physics Laboratory, 
Helsinki University of Technology, 02015 HUT, Finland\\
$^2$Institut f\"ur Theoretische Festk\"orperphysik, Universit\"at Karlsruhe,
76128 Karlsruhe, Germany\\
$^3$Forschungszentrum Karlsruhe, Institut f\"ur Nanotechnologie, Germany}

\date{\today}
\maketitle

\begin{abstract}

In this work, we study the spin dynamics of ${\rm Mn}_{12}$-acetate
molecules in the regime of thermally assisted tunneling.
In particular, we describe the system in the presence of a strong 
transverse  magnetic field.
Similar to recent experiments, the relaxation time/rate is found to 
display a series of resonances; their Lorentzian shape is found to stem 
from the tunneling.
The dynamic susceptibility $\chi(\omega)$ is calculated starting from the 
microscopic Hamiltonian and the resonant structure manifests itself also in 
$\chi(\omega)$.
Similar to recent results reported on another molecular magnet, ${\rm Fe}_8$,
we find oscillations of the relaxation rate as a function of the transverse
magnetic field when the field is directed along a hard axis of the molecules.
This phenomenon is attributed to the interference of the geometrical or 
Berry phase.
We propose susceptibility experiments to be carried out for strong 
transverse magnetic fields to study of these oscillations and
for a better resolution of the sharp satellite peaks in the relaxation rates.

\end{abstract}
\pacs{75.45.+j, 75.50.Xx, 75.40.Gb}
]

\section{Introduction}

In recent years, numerous experimental results on macroscopic samples of
molecular magnets, especially Mn$_{12}$-acetate and Fe$_8$-triazacyclononane, 
have drawn attention to the peculiar resonant structure observed in the 
hysteresis loops and relaxation time measurements,
\cite{Friedmanetal1,Hernandezetal1,Thomasetal1,Hernandezetal2} 
as well as in the dynamic susceptibility.
\cite{Hernandezetal1,Luisetal1}
In this paper, we concentrate on Mn$_{12}$ (shorthand for Mn$_{12}$-acetate).
At low temperature, the observed relaxation times $\tau$ are long, 
up to several months and more,
and display a series of resonances with faster relaxation
as a function of an external magnetic field directed along the easy ($z$)
axis of the sample. 
These are considered as signs of macroscopic quantum tunneling (MQT)
of magnetization.

\begin{figure}
\epsfysize=5.0cm
\centerline{\epsffile{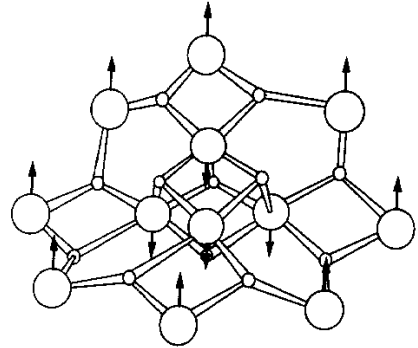}}
\caption{
  Mn$_{12}$-acetate molecule. The big circles denote manganese and 
  the smaller ones oxygen ions. The arrows correspond to the orientation 
  of the atomary spins in the $S=10$ ground state. The figure is reproduced
  from 
  [10] 
  under the copyright of EDP Sciences.
        }
\label{fg:molecule}
\end{figure}

Typical experimental samples consist of single crystals or 
ensembles of aligned crystallites of identical Mn$_{12}$-molecules. 
Each molecule has eight Mn$^{3+}$ and four Mn$^{4+}$ ions 
which, in their ferrimagnetic ground state, have a total spin $S=10$,
see Fig.~\ref{fg:molecule}. 
Due to strong anisotropy along one of the crystalline axes ($z$-direction), 
there is a high potential barrier 
\begin{eqnarray}
  U(S_z)=-AS_z^2-BS_z^4,
\label{eq:barrier}
\end{eqnarray}
with $A/k_{\rm B}\approx0.54$K and $B/k_{\rm B}\approx0.0011$K,
between the opposite orientations of the spin ($S_z=\pm10$); 
the easy axis is the same for all the molecules.
\cite{structure}
The dipolar interaction between the molecular spins, a possible relaxation 
mechanism, has been found to be weak in Mn$_{12}$.
\cite{dipolar1,dipolar2}
Instead, the observed resonant phenomena are attributed to quantum 
{\it tunneling} of single spins -- the response being magnified by 
the large number of them -- interacting with the phonons in the lattice.
The role of the hyperfine interactions is still under some controversy and
is only briefly touched upon in the following.

The main features of the experimental findings can be understood in terms 
of two competing relaxation mechanisms: quantum mechanical
tunneling {\it through} and thermal activation {\it over} the anisotropy 
barrier.
At high temperatures ($T>3{\rm K}$ or $T>6{\rm K}$, depending on the 
experiment
\cite{is_this_so}), 
the spins relax predominantly
via thermal activation due to the phonons in the lattice.
\cite{Villainetal,Villainetal2,Hartmanetal}
In this regime, the relaxation time follows
the Arrhenius law $\tau=\tau_0 \exp(U/k_{\rm B}T)$, where 
$U/k_{\rm B}\approx 60{\rm K}$ denotes the barrier height and 
$\tau_0\approx10^{-8}{\rm s}$ the inverse attempt frequency.
\cite{Arrhenius}
When temperature is lowered to $2{\rm K}<T<3{\rm K}$, the time required 
by the over-barrier relaxation increases exponentially but several of 
the excited states still remain thermally populated. 
In this regime, pairs of states on the opposite sides of the barrier
can be brought to degeneracy by tuning the external magnetic field.
This enhances the probability to tunnel across the barrier; 
the tunneling arises due to crystalline anisotropy and possible transverse
magnetic field at the site of the spin. The tunneling amplitudes
are the larger the closer to the top of the barrier the states are
and, consequently, the thermal population of the higher 
states plays a key role in relaxation, cf. e.g. Refs.
\onlinecite{Villainetal,Villainetal2,Hartmanetal} and 
\onlinecite{Garanin-Chudnovsky}.
At still lower temperatures, tunneling and the relaxation becomes sensitive 
to fluctuations in the dipolar and hyperfine fields.
\cite{Thomasetal2}
In this paper, we concentrate in the regime of thermally-activated tunneling.

Several authors have investigated the spin dynamics theoretically 
with the emphasis ranging from ``minimal'' models, assuming as simple 
a spin Hamiltonian ${\mathcal H}_S$ and a model of the surroundings as 
possible (in order to explain experiments, that is),
\cite{Villainetal,Villainetal2,Hartmanetal,Garanin-Chudnovsky,Luisetal2,Fortetal,LL}
to more specific models for investigating the role of the dipolar and/or 
hyperfine interactions,
\cite{Prokofev-Stamp,Garaninetal,Alsaqeretal}
and combinations of these
\cite{Hartmanetal,Garanin-Chudnovsky,Luisetal2,review}.
The thermally activated relaxation has typically been studied using a master 
equation approach to describe the time evolution of the spin density matrix.
\cite{Villainetal,Villainetal2,Hartmanetal,Garanin-Chudnovsky,Luisetal2,Fortetal,LL,us}
The susceptibility, on the other hand, has only been treated within 
a phenomenological model.
\cite{Luisetal1,Chudnovsky_book}

The existing theories have been successfull in explaining the general 
features seen in experiments. However, several points call for further 
attention.
1) A microscopic calculation of the dynamic susceptibility is missing 
altogether. 
2) The effect of a strong transverse magnetic field has not been 
thoroughly studied and, in particular, not in the context of 
the susceptibility. 
3) Several authors including ourselves have found a series of side 
resonances to arise in their calculations
\cite{Luisetal2,LL,us} 
- the fact that these peaks are not in general
(see Ref.~\onlinecite{Sarachik-Zhong} for exceptions) 
observed in experiments is not quite clear.
In this work, we aim at elucidating these points and
present calculations for the relaxation rates and susceptibility in 
a unified language that can be conveniently extended to systems with 
stronger couplings. 
We work with a Hamiltonian similar to, e.g., Ref.~\onlinecite{LL},
cf. Eqs.~(\ref{eq:Hz}), (\ref{eq:HT}), and (\ref{eq:Hsp}), but introduce 
an alternative framework to work with the density matrix.
It is well known that all the resonances can be enhanced by a strong 
transverse magnetic field but we show that the resonances can also be
reduced and even suppressed:
Both the relaxation rate, 
%
%
see also Ref.~\onlinecite{review}, 
and susceptibility are found to display
significant dependence on the direction of the transverse field
suggesting that interference effects of a geometrical phase could be 
observed also in Mn$_{12}$ and, what is more, do so in the regime of 
thermally-activated tunneling.

The paper is organized as follows.
Part \ref{sec:system} introduces the microscopic model used for Mn$_{12}$ and
a discussion on the different interaction mechanisms in the system.
In part \ref{sec:dynamics}, we develop a time-dependent description of
the system in terms of a real-time diagrammatic technique.
This approach is then used to derive and evaluate the kinetic equation
and the resulting master equation governing the spin dynamics.
In part \ref{sec:kinetic}, we solve the kinetic equation for the 
field-dependent relaxation times $\tau(H)$ and the static susceptibility 
$\chi_0(H)$. The dynamic susceptibility $\chi(\omega;H)$ is calculated in 
part \ref{sec:chi} and a Kubo-type formula is found.
Part \ref{sec:results} displays the numerical results for both $\tau(H)$
and $\chi(\omega;H)$ accompanied with a discussion on the results and their
relevance to experiments. In part \ref{sec:conclusions} we sum up the work.

\section{System}
\label{sec:system}

\begin{figure}
\epsfysize=6.5cm
\centerline{\epsffile{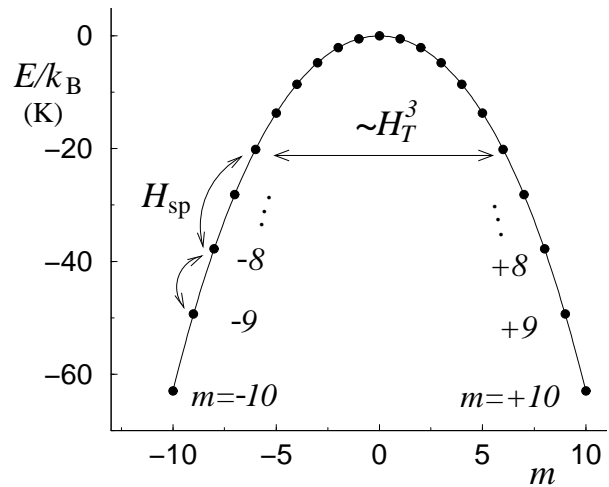}}
\caption{
  Schematic of the energy diagram for $2S+1=21$ eigenstates of 
  ${\mathcal H}_z$ together with the functional form of the potential 
  barrier (solid line).
  The figure also shows examples of possible transition processes: 
  spin-phonon coupling may change $m$ by $\pm1$ or $\pm2$, while 
  the anisotropy term in ${\mathcal H}_{\rm T}$ yields tunnel couplings 
  across the barrier -- an example of a third-order (in $B_4$) process is
  depicted in the figure.}
\label{fg:energies}
\end{figure}

The spin Hamiltonian of a single Mn$_{12}$ molecule 
can be written in the form 
${\mathcal H}_S={\mathcal H}_z+{\mathcal H}_{\rm T}$.
The first term,
\begin{eqnarray}
        {\mathcal H}_z=-AS_z^2-BS_z^4-g\mu_{\rm B}H_z S_z,
\label{eq:Hz}
\end{eqnarray}
with $S_z$ being the spin component along the easy axis 
(here the $z$-direction),
describes the part that commutes with $S_z$. It consists of  
the anisotropy terms of Eq.~(\ref{eq:barrier}) and a Zeeman term which 
enables external biasing of the energies. 
The anisotropy constants have been experimentally estimated
\cite{Barraetal,Zhongetal}
as $A/k_{\rm B}=0.52-0.56{\rm K}$ and $B/k_{\rm B}=0.0011-0.0013{\rm K}$;
the g-factor is 1.9.
\cite{Sessolietal}
The resulting energy levels $E_m$ for the eigenstates of 
$S_z|m\rangle=m|m\rangle$ together with the potential barrier are shown 
schematically in Fig.~\ref{fg:energies}.

The second term in the Hamiltonian,
\begin{eqnarray}
        {\mathcal H}_{\rm T}=-\frac{1}{2}B_4\left(S_+^4+S_-^4\right)
                            -g\mu_{\rm B}(H_xS_x+H_yS_y)
\label{eq:HT}
\end{eqnarray}
does not commute with $S_z$ and gives rise to tunneling.
The $B_4$-term arises from crystalline anisotropy,
$B_4=(4.3-14.4)\cdot10^{-5}{\rm K}$ 
(below we use $B_4=8.6\cdot10^{-5}$K, but the particular choice is
unimportant for the results obtained),\cite{Barraetal} while the second 
term is the Zeeman term corresponding to a transverse magnetic field
$H_\perp=H\sin\theta$ (in spherical coordinates, $\theta$ is the polar angle 
away from the $z$-axis; the azimuth angle is denoted $\phi$:
$H_x=H_\perp\cos\phi$ and $H_y=H_\perp\sin\phi$).

\begin{figure}
\epsfxsize=8.5cm
\centerline{\epsffile{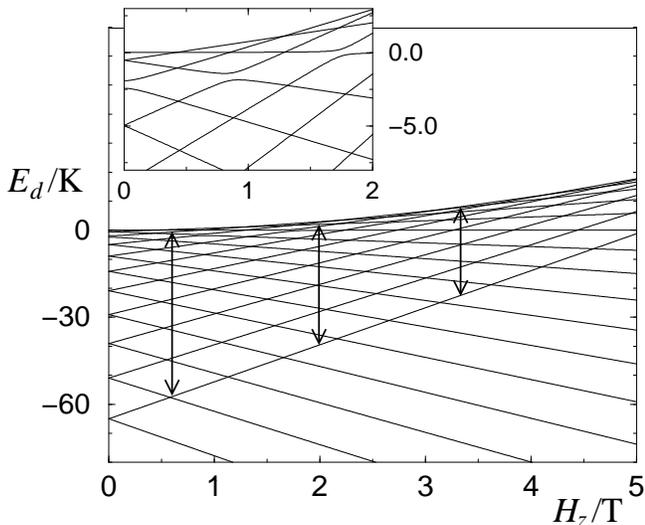}}
\caption{
  The eigenenergies of the 21 states as a function of $H_z$. The inset shows
  a blow up of the higher energies at low fields -- note especially the avoided
  crossings of the states $m-m^\prime=4$ directly coupled by the $B_4$-term 
  in ${\mathcal H}_{\rm T}$. There are similar, although smaller, splittings
  for the lower levels as well.
  The three arrows denote the effective barrier height that
  decreases with increasing $H_z$.
        }
\label{fg:energies_wis}
\end{figure}

Figure \ref{fg:energies_wis} shows the eigenenergies of ${\mathcal H}_S$
as a function of the longitudinal magnetic field $H_z$.
Away from the resonances, the eigenstates $|d\rangle$ resemble the states
$|m\rangle$ and are also localized onto the different sides of the barrier --
the linear field dependence of the eigenenergies stems from 
the Zeeman term in Eq.~(\ref{eq:Hz}).
Close to the resonances, ${\mathcal H}_{\rm T}$ couples the $|m\rangle$
states across the barrier and gives rise to avoided crossings in the energy 
diagram for $E_d$.
The magnitude of these splittings directly gives the tunneling strengths.
Depending on the states in question as well as on the magnitude of $B_4$
($H_\perp=0$ for the moment), 
the splittings are found to vary enormously: from $10^{-10}$K 
for the choice $B_4/k_{\rm B}=4.3\cdot10^{-5}$K and the states $m=\pm10$
up to almost 2K for $B_4/k_{\rm B}=14.4\cdot10^{-5}$K and the resonance 
$m=\pm2$.
This upper limit is already of the same order of magnitude as the level
spacing and in fact even exceeds it rendering perturbative calculations of 
the tunneling couplings/strengths somewhat questionable.
Therefore, even though we first formulate everything in a general form, 
independent of the choice of basis for the spins, we calculate
the actual results working in the eigenbasis of ${\mathcal H}_S$.
This choice of basis provides the additional advantage that it allows us 
to consider arbitrarily strong magnetic fields.

\begin{figure}
\epsfxsize=8.5cm
\centerline{\epsffile{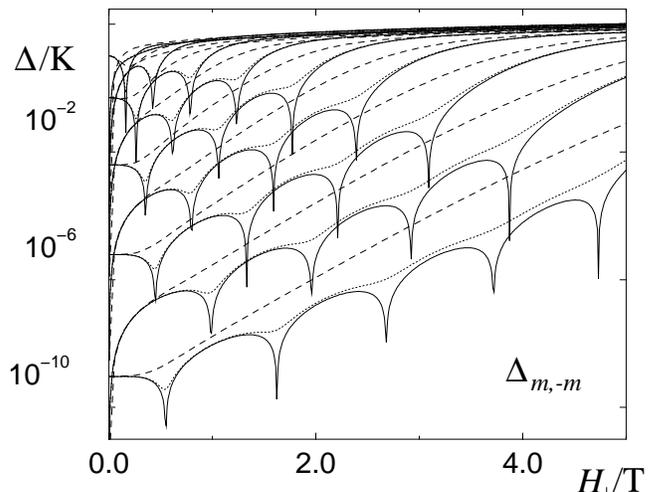}}
\caption{
  The tunnel splittings $\Delta_{m,-m}$ as a function of the
  transverse magnetic field; $H_z=0$. 
  The three types of curves correspond to
  three different angles $\phi$: $0^\circ$ -- dashed, 
  $40^\circ$ -- dotted, and $45^\circ$ -- solid.
        }
\label{fg:splittings}
\end{figure}

In absence of $H_\perp$, there is a selection rule to ${\mathcal H}_{\rm T}$:
only states $m$ and $m^\prime$ 
that are a multiple of four apart
are coupled. Experimental data do not lend support to such a rule,
however, but rather suggest that all transitions are allowed.
It turns out that already a tiny transverse field in Eq.~(\ref{eq:HT})
is sufficient in achieving this
-- such a field may arise due to dipolar and/or hyperfine interactions
within the sample, see below, as well as due to uncertainity in 
the precise angle between the external field and the easy 
axis of the sample. 
We assume throughout the paper 
a small constant misalignment angle $\theta=1^\circ$ and 
$H_\perp=|{\it\bf H}|\cdot\sin\theta$.
In places, we wish to investigate the effects of a significantly 
stronger transverse field and state so explicitly.
The transverse field increases the tunnel splittings 
if $\phi$ is close to $n\pi/2$ ($n$ is an integer), i.e., 
along the $x$ and $y$-axes,
or leads to oscillations in the splittings if $\phi$ is close to one
of the directions $\pi(2n+1)/4$.
Below we denote these special directions as the hard axes of the molecules.
\cite{fe8}
The tunnel splittings are illustrated in Fig.~\ref{fg:splittings} 
as a function of $H_\perp$ and for different angles $\phi$,
%
%
see Ref.~\onlinecite{berries}.

The usage of ${\mathcal H}_S$ of an isolated spin is based, first of all, on 
the assumption that the molecules indeed reside in their $S=10$ ground state.
\cite{comment_on_Alsaqer}
This is well justified for the experimental temperatures below $3-6{\rm K}$ 
-- the energy required to excite the system to the lowest excited state with 
$S=9$ is around 30K. 
The second assumption is the absence of interaction.
In reality, the spins interact with each other via dipolar interaction, 
with the nuclear spins via hyperfine interaction, and with the phonons of 
the surrounding lattice. Experimental evidence shows that the dipolar 
interactions are small for $T>2-2.4{\rm K}$,
\cite{Thomasetal2,Wernsdorfer_EPL}  
while the hyperfine interactions produce an intrinsic broadening 
of the order of 10mT to the spin states.
\cite{Hartmanetal,Wernsdorfer_EPL}
We neglect these for the moment and return to them in 
Sec.~\ref{sec:discussion}.

The spin-phonon interaction is mediated by variations in the local 
magnetic field induced by lattice vibrations and distortions. 
For tetragonal symmetry, this can be generally formulated as,
cf. Ref.~\onlinecite{LL},
\begin{eqnarray}
\label{eq:Hsp}
  {\mathcal H}_{\rm sp} & = & 
        g_1(\epsilon_{xx}-\epsilon_{yy})\otimes\left(S_x^2-S_y^2\right)
        +\frac{1}{2} g_2 \epsilon_{xy}\otimes\left\{S_x,S_y\right\} \nonumber\\
        &+&\frac{1}{2} g_3 \left(\epsilon_{xz}\otimes\left\{S_x,S_z\right\}
                        +\epsilon_{yz}\otimes\left\{S_y,S_z\right\}\right)\\
        &+&\frac{1}{2} g_4 \left(\omega_{xz}\otimes\left\{S_x,S_z\right\}
                  +\omega_{yz}\otimes\left\{S_y,S_z\right\}\right)\nonumber
\end{eqnarray}
where $\epsilon_{\alpha\beta}$ and $\omega_{\alpha\beta}$ are 
the symmetric and antisymmetric strain tensors, respectively, and
$g_i$ ($i$=1,2,3,4) are the spin-phonon coupling constants.
For these, we adopt the values from Ref.~\onlinecite{LL}:
$g_1=g_4/2=A$ and $|g_2|\approx g_1$ and $|g_3|\approx g_4$.
In leading order in $g_i$'s, ${\mathcal H}_{\rm sp}$ produces transitions 
such that for, $g_{1,2}$-terms, $\Delta m=\pm2$ and, for $g_{3,4}$-terms, 
$\Delta m=\pm1$. 
The phonons themselves are described by 
\begin{eqnarray}
  {\mathcal H}_{\rm ph}=\sum_{\vec{k}\sigma}
       \omega_{\vec{k}\sigma} b_{\vec{k}\sigma}^\dagger b_{\vec{k}\sigma}^{}
\end{eqnarray}
as a bath of noninteracting bosons.

To be more quantitative, the phonons are assumed to be plane waves with 
a linear spectrum and three modes -- two transverse and one longitudinal --
denoted by $\sigma$. The elements of the strain tensor, 
$\epsilon_{\alpha\beta}\equiv(\partial_\alpha u_\beta + \partial_\beta u_\alpha)/2$
and $\omega_{\alpha\beta}\equiv(\partial_\alpha u_\beta - \partial_\beta u_\alpha)/2$,
are defined in terms of the local displacement vector
\begin{eqnarray}
  u_\alpha(\vec{r}) &=& \sum_{\vec{k}\sigma}
                \sqrt{\frac{\hbar}{2NM\;\omega_{\vec{k}\sigma}}}
                                                \;e_\alpha^{(\sigma)}\;
                [b_{\vec{k}\sigma}^\dagger + b_{\vec{k}\sigma}^{}]
                                                    \;e^{i\vec{k}\cdot\vec{r}}.
\label{eq:displacement}
\end{eqnarray}
Here $b_{\vec{k}\sigma}^{(\dagger)}$ are the bosonic operators for phonons
with wave vector $\vec{k}$, $\omega_{\vec{k}\sigma}$ is the correponding 
frequency, and $e_\alpha^{(\sigma)}$ the $\alpha$th element (of $x$, $y$, 
and $z$) of the polarization vector;
$N$ is the number of unit cells and $M$ the mass per unit cell.

Above we considered the energy scales inherent to ${\mathcal H}_S$, and
the spin-phonon rates, which are found below (Apps.~\ref{app:rules} and 
\ref{app:sigma}) to be typically of the order 
of $10^{-5}-10^{-4}$K, fall in between the extremes of the tunnel splittings.
This value is very small compared to the stronger tunneling couplings
and seems to suit well for a perturbative treatment. 
On the other hand, for the low-lying and weakly coupled states
the spin-phonon rates may be several orders of magnitude larger than the 
tunnel splittings and one would expect the tunneling to be suppressed.
However, the intermediate regime, where the tunneling and spin-phonon
rates are of the same order of magnitude, requires some extra care,
see below.

\section{Dynamics}
\label{sec:dynamics}

The magnetization measured in experiments is the molecular magnetization 
$M(t)$ magnified by the large number of molecules in the samples.
Let us define $M(t)$ in terms of the reduced density matrix
$\rho(t)= {\rm Tr}_{\rm ph}[\rho^{\rm tot}(t)]$ ($\rho^{\rm tot}(t)$
is the full density matrix)
\begin{eqnarray}
\label{eq:mtp}
  M(t)   &=& g\,\mu_{\rm B} \; \langle S_z(t) \rangle\\
         &\equiv& g\,\mu_{\rm B} \; {\rm Tr}_S [S_z\rho(t)]
         = g\,\mu_{\rm B}\sum_m m\cdot \rho_{m,m}(t)\nonumber
\end{eqnarray}
The reduced density matrix $\rho(t)$
describes the spin degrees of freedom in the presence of
the phonon reservoir -- its diagonal elements are just the 
probabilities for the spin to be in the states $|m\rangle$. 
Our strategy is to start with the general formulation 
\begin{eqnarray}
  \langle S_z(t) \rangle = {\rm Tr}[S_z(t)\cdot\rho_0^{\rm tot}],
\label{eq:magnetization}
\end{eqnarray}
where $\rho_0^{\rm tot}=\rho^S\rho^{\rm ph}$ denotes the initial density 
matrix of the whole system encompassing the spin and phonon degrees of freedom.
However, this does not contain the interaction between the two: we assume that
the coupling is turned on adiabatically and only enters the time evolution of 
$S_z(t)$ (in Heisenberg picture and with the convention $\hbar=1$)
\begin{eqnarray}
  \langle S_z(t)\rangle = {\rm Tr}[
        e^{+i\!\int_{t_0}^t\!\!dt^\prime {\mathcal H}(t^\prime)} S_z \;
           e^{-i\!\int_{t_0}^t\!\!dt^\prime {\mathcal H}(t^\prime)}
        \cdot\rho_0^{\rm tot}].
\label{eq:expression1}
\end{eqnarray}

In order to evaluate Eq.~(\ref{eq:expression1}) and to describe 
the dynamics of the system more quantitatively,
we introduce a real-time diagrammatic language that is applicable to 
any mesoscopic system comprising a part with a finite number of states 
linearly coupled to an external heat (or particle) reservoir.
\cite{diagrams}

\subsection{Diagrammatic language}
\label{sec:diagrams}

\begin{figure}
\epsfxsize=7.0cm
\centerline{\epsffile{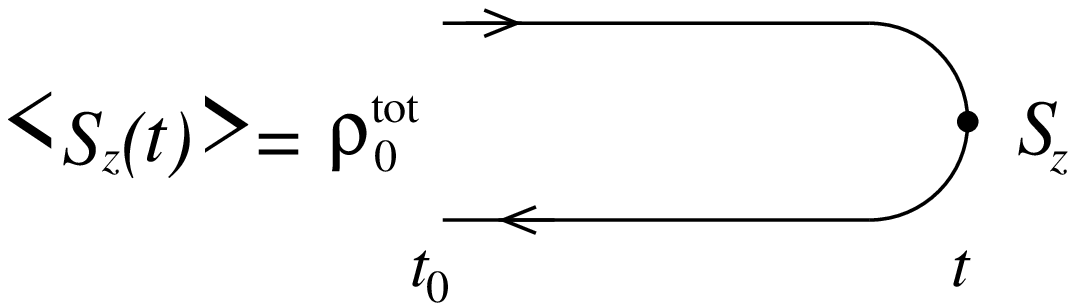}}
\caption{
        Diagrammatic representation of Eq.~(\ref{eq:expression1}),
        reading the expression from right to left and starting
        at time $t_0$ at the initial state described by 
        $\rho_0^{\rm tot}$, then propagating forward (upper propagator)
        in time to time $t$, where $S_z$ (vertex) is to be evaluated, and 
        returning backwards (lower propagator) in time to $t_0$.
        In such diagrams, the trace is always implicitly assumed.
        }
\label{fg:diagram1}
\end{figure}

Equation (\ref{eq:expression1}) can be written as a diagram depicted 
in Fig.~\ref{fg:diagram1}.
In this equation, the spin-phonon interaction terms can be separated 
from the noninteracting part of the Hamiltonian by shifting to the 
interaction picture with respect 
to ${\mathcal H}_0={\mathcal H_S}+{\mathcal H}_{\rm ph}$
(indicated with the subscript I)
\begin{eqnarray}
\label{eq:interaction}
   \langle S_z(t) \rangle &=& {\rm Tr} [\rho_0^{\rm tot}\cdot S_z(t)]\\[5pt]
        &=& {\rm Tr} [ \rho_0^{\rm tot} \cdot
        \tilde{T} e^{+i\!\int_{t_0}^t\!\!dt^\prime 
                                {\mathcal H}_{\rm sp}(t^\prime)_{\rm I}}
        S_{z}(t)_{\rm I}\;
                T e^{-i\!\int_{t_0}^t\!\!dt^\prime 
                        {\mathcal H}_{\rm sp}(t^\prime)_{\rm I}}]\nonumber
\end{eqnarray}
where ($\tilde{T}$) $T$ is the (anti-)time-ordering operator and 
$\rho_0^{\rm tot}$ has been moved to the front 
(this is allowed by the invariance of trace under a cyclical permutation of 
its arguments).

\begin{figure}
\epsfxsize=6.0cm
\centerline{\epsffile{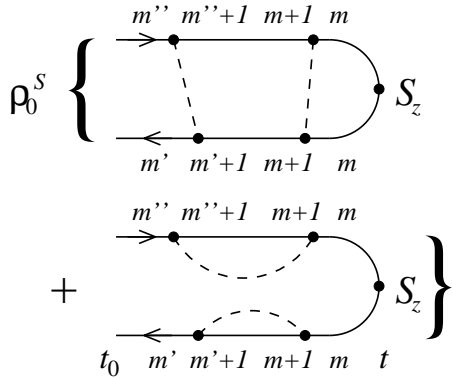}}
\caption{
        Examples of the diagrams arising 
        after integration over phonons. The indices on the forward (backward)
        propagators correspond to the spin states and, in particular, 
        the first (second) index of $\rho(t)$. 
        Note that in general $m$, $m^\prime$, and $m^{\prime\prime}$ 
        may all be unequal if tunneling is allowed along the propagator
        and if we also consider the nondiagonal elements of $\rho(t)$.
        }
\label{fg:diagram3}
\end{figure}

We proceed by eliminating the phonon degrees of freedom with
the aim to obtain an effective theory for the dynamics of the
spin system. We first separate the trace and introduce time-ordering
$T_{\rm K}$ along the closed Keldysh-contour in Fig.~\ref{fg:diagram1}
\begin{eqnarray}
  {\rm Tr}[\rho_0^{\rm tot}\!\!&\!\cdot\!&\!\!S_z(t)] = 
        \nonumber\\[2pt]
\label{eq:Keldysh}
   &=& {\rm Tr}_S\{\rho_0^S \; {\rm Tr_{ph}}\rho_0^{\rm ph}\;
  T_{\rm K}e^{-i\int_{\rm K} dt^\prime {\mathcal H}_{\rm sp}(t^\prime)_{\rm I}}
        S_z(t)_{\rm I}\}.
\label{eq:reduced}
\end{eqnarray}
The exponential, expanded  
in powers of ${\mathcal H}_{\rm sp}$, reads
\begin{eqnarray}
\label{eq:expansion}
&&T_{\rm K}e^{-i\int_{\rm K} dt^\prime {\mathcal H}_{\rm sp}(t^\prime)_{\rm I}} = \sum_{m=0}^\infty (-i)^m\\[-2pt]
&& \;\;\cdot\!\int_{\rm K}\!dt_1\!\int_{\rm K}\!dt_2...\!\int_{\rm K}\!dt_m
T_{\rm K}\{{\mathcal H}_{\rm sp}(t_1)_{\rm I}\;{\mathcal H}_{\rm sp}(t_2)_{\rm I}\;...\;{\mathcal H}_{\rm sp}(t_m)_{\rm I}\}\nonumber
\end{eqnarray}
with $t_1>t_2>...>t_m$. 
Each ${\mathcal H}_{\rm sp}(t_i)_{\rm I}$ is represented as a vertex
on the Keldysh contour. 
In the next step, this expansion together with the explicit expression
(\ref{eq:Hsp}) for ${\mathcal H}_{\rm sp}$ is inserted into 
Eq.~(\ref{eq:reduced}) and
the trace over the phonons is performed by using Wick's theorem. 
As a consequence, the phonon operators are pairwise contracted:
${\rm Tr}_{\rm ph}\rho_0^{\rm ph}b^\dagger(t)b(t^\prime)=
\langle b^\dagger(t)b(t^\prime) \rangle$ and 
${\rm Tr}_{\rm ph}\rho_0^{\rm ph}b(t)b^\dagger(t^\prime)=
\langle b(t)b^\dagger(t^\prime) \rangle$. 
Here, the phonons are considered as a reservoir that is 
not perturbed by the single spin -- the bath remains in equilibrium 
and the contractions are given by the Bose-Einstein distribution 
$\langle b_{\vec{k}\sigma}^\dagger b_{\vec{k}\sigma}^{}\rangle=
[\exp(\beta \omega_{\vec{k}\sigma})-1]^{-1}$.
In terms of the diagrams this means that the vertices get coupled 
(in all possible ways) by {\it interaction} or {\it phonon lines} 
which correspond to just these contractions, cf. Fig.~\ref{fg:diagram3}.
The general rules for evaluating the diagrams are given in 
App.~\ref{app:rules} and the contributions from the interaction lines
are calculated in App.~\ref{app:sigma}.

\begin{figure}
\epsfxsize=8.5cm
\centerline{\epsffile{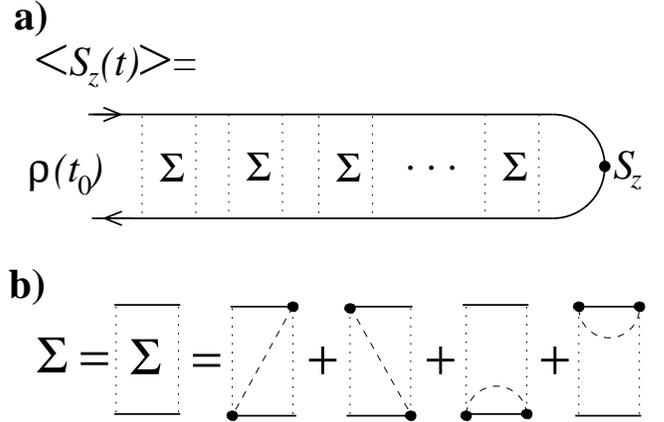}}
\caption{
  Diagrammatic expression for (a) the expectation value of $S_z$ in terms
  of the element $\Sigma$, which corresponds to the phonon-induced 
  transition rate of the spin. (b) Diagrammatic representation for
  $\Sigma$ in $O(g^2)$.   
        }
\label{fg:S_and_Sigma}
\end{figure}

All the diagrams, e.g., those in Fig.~\ref{fg:diagram3}, are composed of
two kinds of elements that can be distinguished by drawing a vertical line 
through the diagram and checking whether it cuts phonon lines or not. 
If it {\it does not}, the corresponding part of the diagram represents
free evolution of the spin; if it {\it does}, the part of the diagram between
successive periods of free evolution corresponds to spin-phonon interaction
and, more particularly, to a transition rate between different spin states.
The sum of all the possible diagrams with the interaction lines
is denoted as $\Sigma$ and, when evaluated, its terms are $O(g^{2n})$ 
where $g$ is the spin-phonon coupling constant and $n$ the number of
interaction lines in the diagram,
cf. Appendices \ref{app:rules} and \ref{app:sigma}, and below.
The full time evolution of $\langle S_z(t) \rangle$ can be expressed as in 
Fig.~\ref{fg:S_and_Sigma}a in terms of the two kinds of contributions, 
just discussed.
In lowest order in $g$, $\Sigma$ is given by the diagrams shown in part (b) 
of the figure.

\subsection{Kinetic equation}
\label{sec:kinetic}

The reduced density matrix defined as
\begin{eqnarray}
  \rho_{m,m^\prime}(t)\equiv 
        \big\langle (|m^\prime\rangle\langle m|)(t)\big\rangle
\label{eq:reduced2}
\end{eqnarray}
accounts for the full time evolution of the spin in the presence of 
the spin-phonon interaction. Its diagrammatic expansion is analog
to Fig.~\ref{fg:S_and_Sigma}a with $S_z$ being replaced by
$|m^\prime\rangle\langle m|$. Setting $t_0=0$ and differentiating
with respect to time, we obtain the kinetic equation
\begin{eqnarray}
  \dot{\rho}(t)+i[{\mathcal H}_S,\rho(t)]=\int_0^t dt^\prime
        \Sigma(t,t^\prime)\rho(t^\prime)
\label{eq:kinetic}
\end{eqnarray}
The commutator on the l.h.s. of Eq.~(\ref{eq:kinetic}) corresponds to 
the free (Hamiltonian) time evolution of the spin, while the integral 
on the r.h.s. describes a dissipative interaction which is nonlocal 
in time and contains the spin-phonon coupling terms.

The time dependence of the kernel $\Sigma(t,t^\prime)=\Sigma(t-t^\prime)$ 
is evaluated explicitly in Apps.~\ref{app:rules} and \ref{app:sigma}
up to order $O(g^2)$, 
compare  Fig.~\ref{fg:S_and_Sigma}b. It depends only on the time difference
since the Hamiltonian is time-translationally invariant.
According to Eq.~(\ref{eq:correlator_t}), we find that $\Sigma(t-t^\prime)$
is a fast decaying function of time, and we make the 
simplifying Markov assumption that $\rho(t)$ remains essentially 
constant over the time period $\Delta t$ within which $\Sigma(t-t^\prime)$ 
decays to zero and take $\rho(t)$ in front of the integral.
\cite{Markov} This is justified at least for the longest relaxation time
$\tau_1$ (see below).
For convenience, we also take the upper integration limit to infinity.

After taking $\rho(t)$ out of the integral in Eq.~(\ref{eq:kinetic})
the integration over time (up to infinity) can be performed and we are
left with a constant $\Sigma$. This is evaluated in App.~\ref{app:sigma}
and corresponds to the diagrams in Figs.~\ref{fg:S_and_Sigma}b and 
\ref{fg:diagram_pairs}a and b, see also below.
The kinetic equation (\ref{eq:kinetic}) becomes
\begin{eqnarray}
  \dot{\rho}(t)=-iL_0\,\rho(t)+\Sigma\,\rho(t)\equiv W\rho(t)
\label{eq:kinetic_W}
\end{eqnarray}
with $L_0\equiv[{\mathcal H}_S,.]$. 
This is similar to the master equation formulations in 
the literature
\cite{Hartmanetal,Garanin-Chudnovsky,Luisetal2,Fortetal,LL}
and may be solved for the eigensolutions of $W$
\begin{eqnarray}
  \dot{\rho}^{(i)}(t)&=&W\rho^{(i)}(t)=\lambda_i\rho^{(i)}(t)\\
  \rho^{(i)}(t)&=&\rho^{(i)}(0)\cdot e^{\lambda_i t}.
\label{eq:eigenvalues}
\end{eqnarray}
The real parts of the eigenvalues correspond to relaxation rates 
$Re(\lambda_i)=-\tau_i^{-1}$.

The eigensolution of $W$ with $\lambda_0=0$ corresponds to the stationary 
state, defined as $\dot{\rho}^{(0)}(t)=0$ with $\rho^{(0)}(t)=\rho^{(0)}$. 
The diagonal elements of $\rho$ are the thermal probabilities to find 
the spin in the respective states.
In the $m$-basis, the static magnetization and susceptibility are
readily expressed in terms of the diagonal components 
$\rho_m^{(0)}=\rho_{m,m}^{(0)}$ yielding
\begin{eqnarray}
\label{eq:M_0}
  M_0 = g\mu_{\rm B}\sum_m m\,\rho_m^{(0)} 
        \!\!\!\!\!\!\!\!\!\!\!\!\!\!\!\!\!\!\!\!\!\!\!&&\\
  \chi_0 = \frac{\partial M_0}{\partial H_z} &=& 
        g\mu_{\rm B}\sum_m m\cdot\frac{\partial \rho_m^{(0)}}{\partial H_z}.
\end{eqnarray}
Since the stationary probabilities are given by Boltzmann factors we obtain
in the absence of tunneling
\begin{equation}
\label{eq:chi_0}
\chi_0= -\frac{(g\mu_{\rm B})^2}{k_{\rm B}T}
   \sum_{m,m^\prime}m(m^\prime-m)\rho_m^{(0)}\rho_{m^\prime}^{(0)}.
\end{equation}
In the precence of tunneling, this form remains essentially the same
up to transverse fields of the order of $1{\rm T}$ (in this range most of
the tunneling splittings are too small compared to the level spacing in
order to change the result considerably).

For all the other solutions $Re(\lambda_i)<0$ and the vectors $\rho^{(i)}(t)$ 
correspond to deviations from $\rho^{(0)}$. The longest relaxation time 
$\tau_1$ is several orders of magnitude larger than $\tau_2$ and
the respective solution is interpreted to correspond to the over-barrier 
relaxation.

Above we have considered the full (reduced) density matrix $\rho(t)$.
In appendix \ref{app:basis}, we discuss the choice of the basis and 
argue that the $d$-basis has certain advantages and, e.g., in the case
of strong tunneling compard to the spin-phonon rates, it allows the 
restriction to the diagonal elements of 
$\rho(t)_{d,d^\prime}$ only. The nondiagonal elements become important
when the rates for the tunneling and the spin-phonon coupling 
are of the same order of magnitude.

\subsection{Ac-susceptibility}
\label{sec:chi}

In this section we derive an expression for the dynamic susceptibility
\begin{eqnarray}
\label{eq:chi_w}
  \chi(\omega) &\equiv& \int_0^\infty d\tau e^{i\omega \tau}\chi(\tau)\\[5pt]
  \chi(t-t^\prime) &=& 
	\frac{\partial\!\acute{}\; M(t)}{\partial\!\acute{}\; H_z(t^\prime)} 
 = g\,\mu_{\rm B}\frac{
	\partial\!\acute{}\; \langle S_z(t)\rangle}
	{\partial\!\acute{}\; H_z(t^\prime)}
\label{eq:chi_t}
\end{eqnarray}
starting from the Hamiltonian ${\mathcal H}$ and formulating 
the expectation value in Eq.~(\ref{eq:chi_t}) in terms of diagrams.
For convenience, we assume in the following that $H_\perp$ can be tuned to 
any (static) value independent of $H_z$ and that the actual measurement is 
done by applying a tiny ac-excitation field $h_z(\omega)$ on top of 
the static $H_z$. 
The more general calculation of 
$\partial\!\acute{}\; M_\alpha(t)/\partial\!\acute{}\; H_\beta(t^\prime)$, $\alpha,\beta=x,y,z$,
can be carried out along the same lines.

The derivation with respect to $H_z(t^\prime)$ acts on the terms
$\exp[\mp i\int \!dt^\prime{\mathcal H}_z(t^\prime)]\propto\exp[\mp i (-g\,\mu_{\rm B})\int \!dt^\prime S_z(t^\prime)H_z(t^\prime))]$ in Eq.~(\ref{eq:expression1})
and may occur on either the forward or backward propagator of the diagrams
(minus and plus signs, respectively).
The derivation takes down a factor $\pm ig\mu_{\rm B}S_z(t^\prime)$ and
we obtain
\begin{eqnarray}
  \chi(t-t^\prime)= i(g\mu_{\rm B})^2\langle [S_z(t),S_z(t^\prime)] \rangle
\label{eq:Kubo}
\end{eqnarray}
which, when inserted to Eq.~(\ref{eq:chi_w}), is just the Kubo formula 
for the linear response to an external magnetic field.

Susceptibility can be expressed diagrammatically starting from either of 
the equations (\ref{eq:chi_t}) or (\ref{eq:Kubo}). In either case, $S_z(t)$
is written at the latest point to the right.
For Eq.~(\ref{eq:chi_t}), $\partial\!\acute{}\;/\partial\!\acute{}\; H_z(t^\prime)$ may act on 
either of the propagators, while, for Eq.~(\ref{eq:Kubo}), the terms of 
the commutator are ordered on the contour reading them from right (earlier) 
to left (later times).
It is the most convenient to work in the eigenbasis where it is sufficient 
to consider the diagonal elements only, see App.~\ref{app:basis}.
In this case, if $S_z(t^\prime)$
falls in between $\Sigma$'s in the diagram in Fig.~\ref{fg:S_and_Sigma}a,
the commutator equals zero -- we only need to be concerned with the
$\Sigma$ parts.

\begin{figure}
\epsfxsize=8.5cm
\centerline{\epsffile{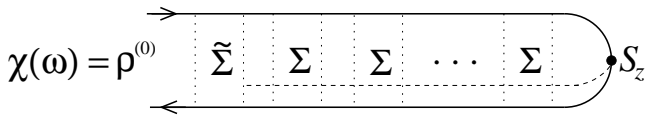}}
\caption{
  Diagrammatic representation of $\chi(\omega)$, see text for details.
        }
\label{fg:chi}
\end{figure}

Let us set $t_0=0$ and define $\tilde{\Sigma}(t^{\prime\prime},0)$ as in 
Fig.~\ref{fg:S_and_Sigma}b but with $S_z(t^\prime)$ 
($0<t^\prime<t^{\prime\prime}$) inserted onto one of the propagators.
Furthermore, let us assign the part $e^{i\omega(t^{\prime\prime}-t^\prime)}$ 
of the Laplace transform to the definition of 
$\tilde{\Sigma}(t^{\prime\prime},0)$.
This is then the part of the integrand in Eq.~(\ref{eq:chi_w})
appearing between times $t=0$ and $t=t^{\prime\prime}$.
On the forward propagator, this yields a factor 
$ig\mu_{\rm B}m_{d,d^\prime} \equiv ig\mu_{\rm B}\langle d^\prime|S_z|d\rangle$
when acting on the state $d$ on the propagator
and changing this to the state $d^\prime$ ($d$-basis is not the eigenbasis
of $S_z$).
On the backward propagator, there is a further minus sign.
As to the whole diagram, see Fig.~\ref{fg:chi}, the system is in 
the stationary state before $t_0=0$, described by $\rho^{(0)}$ --
this corresponds to the requirement that the expectation value 
$\langle S_z\rangle$ is taken with respect to the stationary state.
[One could extend the definition of the susceptibility to account for
experiments with an additional time-dependent (e.g. constantly sweeped)
magnetic field $H_z(t)$ and consider initial states before $t^\prime$ that 
are combinations of the stationary and the relaxing modes.]
After $\tilde{\Sigma}(t^{\prime\prime},0)$, there may be any number of 
$\Sigma$'s between this and the final time $t$ -- the sum of such series 
is denoted $\Pi(t,t^{\prime\prime})$. 
Finally the resulting diagram/expression is Laplace transformed -- 
$e^{i\omega(t-t^\prime)}$ being divided into two parts, one denoted by 
the dashed line running through the diagram from time $t^{\prime\prime}$ 
to $t$ and the other one extending from $t^\prime$ to $t^{\prime\prime}$
and already belonging to the above definition of $\tilde{\Sigma}$.

In evaluating the diagram in Fig.~\ref{fg:chi}, we vary $t^\prime$ within 
$\tilde{\Sigma}(t^{\prime\prime},0)$, and
extend the integrations of $t$ and $t^{\prime\prime}$ to infinity. 
This yields
\begin{eqnarray}
&& \int_{t^\prime}^\infty dt e^{i\omega(t-t^\prime)}
                           \langle[S_z(t),S_z(t^\prime)]\rangle=
	\sum_{d,d^\prime,d^{\prime\prime}}m_d^{}\\
&& \int_0^\infty \!\!\! dt^{\prime\prime} 
        \int_{t^{\prime\prime}}^\infty \!\!\! dt 
        \int_0^{t^{\prime\prime}} \!\!\! dt^\prime
        \Pi(t,t^{\prime\prime})_{d,d^{\prime\prime}}^{}
        \tilde{\Sigma}(t^{\prime\prime},0)_{d^{\prime\prime},d^\prime}^{}
        \rho_{d^\prime}^{(0)}
        e^{i\omega(t-t^{\prime\prime})}\nonumber,
\end{eqnarray}
where, for brevity, we denote the diagonal states by just one index $d$,
e.g., $m_d=m_{d,d}$, see above. 
The integrations can be carried out one by one in the order $t^\prime$, $t$ 
(which yields $\Pi(z=\omega)$), and $t^{\prime\prime}$ to yield
\begin{eqnarray}
\label{eq:chi_exp}
  &&\chi(\omega) = -(g\mu_{\rm B})^2 \\
  &&\;\;\;\;\;\;\;
  \sum_{d,d^\prime,d^{\prime\prime}} \,m_d^{} 
  \left[\frac{1}{-i\omega+iL_0-\Sigma(\omega)}\right]_{d,d^{\prime\prime}}^{}
  \tilde{\Sigma}(\omega)_{d^{\prime\prime},d^\prime}^{}
  \rho^{(0)}_{d^\prime},\nonumber
\end{eqnarray}
where $\Sigma(\omega)$ is defined as the Laplace transform of $\Sigma(t,0)$.
The explicit expressions for $\tilde{\Sigma}(\omega)_{d,d^{\prime\prime}}^{}$ 
turn out lengthy and are omitted here.

The resolvent in Eq.~(\ref{eq:chi_exp}) is treated in terms of 
the eigensolutions of $W\rho(t)$ and the appropriate projections 
of the terms
$\tilde{\Sigma}(\omega)\rho^{(0)}=\sum_i c_i\rho^{(i)}$ are used.
In particular, for the component along the relaxing mode $\rho^{(1)}(t)$, 
the resolvent in Eq.~(\ref{eq:chi_exp}) reads 
\begin{eqnarray}
\label{eq:chi_denom}
  \frac{1}{-i\omega+iL_0-\Sigma(\omega)}
  =\frac{1}{-i\omega-W(\omega)}
  \approx\frac{1}{-i\omega+\tau_1^{-1}}.\nonumber
\end{eqnarray}
The approximate sign is due to the Markov approximation where
$\Sigma(\omega)\approx\Sigma(0)$, cf. App.~\ref{app:Laplace}.
Apart from one $\tau_1$ this is the well-known factor in expressions 
for susceptibility,
see e.g. Refs. \onlinecite{Kubo_book} or \onlinecite{Chudnovsky_book}.
The rest of Eq.~(\ref{eq:chi_exp}) reduces to a prefactor roughly
proportional to $\tau_1^{-1}$ but with weak $H_z$ and
$\omega$-dependences.
For the low frequency limit, we recover the static susceptibility, i.e., 
$\lim_{\omega\rightarrow 0}\chi(\omega)=\chi_0$ where $\chi$ is given by
Eq.~(\ref{eq:chi_0}).

\section{Results}
\label{sec:results}

In this section, we concentrate on results obtained in the eigenbasis
of ${\mathcal H}_S$. This choice is advocated by three points. 
First, it allows us to consider much stronger transverse fields 
(see the discussion on energy scales in Part II)
than the common approach to calculate tunnel splittings in 
the leading-order perturbation theory;
\cite{Garanin,Garanin-Chudnovsky}
second, the origin of the Lorentzian shape of the resonances is seen
to arise naturally from the tunnel splittings.
\cite{Luisetal2,us}
The third point is that, in this basis, all the relevant properties of 
the system are captured by the diagonal elements $\rho(t)_{d,d}$ only.
\cite{full_matrix} 
This feature also improves the performance of the numerics.
We also present results obtained with the full $\rho(t)$ and discuss
the differences as examples of phonon-induced decoherence.
The energies are expressed in kelvin, magnetic fields in teslas, and
rates in $s^{-1}$.

\subsection{Relaxation rates}
\label{sec:results_tau}

\begin{figure}
\epsfxsize=8.5cm
\centerline{\epsffile{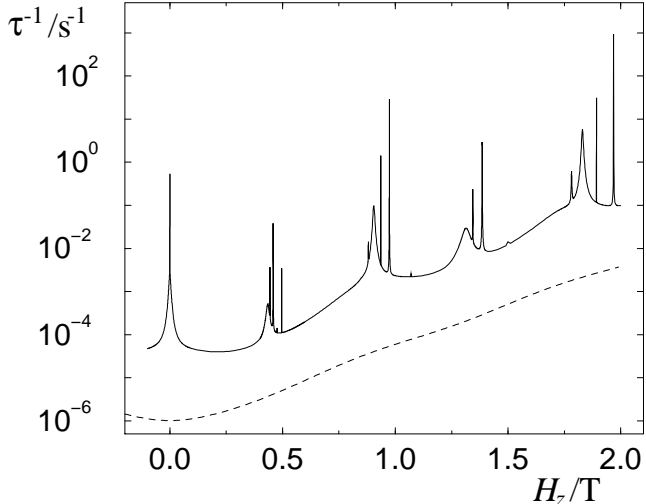}}
\caption{
  Relaxation rate $\Gamma=1/\tau$ as a function of the longitudinal 
  magnetic field: without tunneling (dashed) and with tunneling (solid).
  Here, only the diagonal density matrix elements are used.
  The large offset between the two curves is due to the broad
  Lorentzian shapes under the even resonances: there is a strong
  direct coupling between the states $m=-m^\prime=-2$ around $H_z=0$, 
  $m=-3,m^\prime=1$ under the second series of peaks ($H_z\approx2H_1$), 
  and $m=-4,m^\prime=0$ under the fourth series of peaks ($H_z\approx4H_1$). 
  The sharp resonances arise due to tunnel coupling of
  lower-lying states $m$; their height reflects the energies $E_m$ via 
  the Arrhenius law $\Gamma\propto\exp[-\beta(E_m-E_{-10}])$, and 
  the peak width is given by the tunnel splitting. 
  Here the temperature is 2.5K.
        }
\label{fg:relaxation}
\end{figure}

The relaxation rate as a function of the longitudinal magnetic 
field $H_z$ is shown in Fig.~\ref{fg:relaxation}. The overall behaviour 
is of the Arrhenius form, i.e., $\tau\approx\tau_0\exp[\beta U(H_z)]$
where $\tau_0$ only gives a constant offset to the semilogarithmic figure
and $U$ is the effective height of the barrier, 
see Fig.~\ref{fg:energies_wis}. 
On top of this exponential field dependence, there are series of resonances
located at
\begin{eqnarray}
  H_z^{m,m^\prime}=-\frac{m+m^\prime}{g\mu_{\rm B}}
        \left[A+B(m^2+{m^\prime}^2)\right]
\end{eqnarray}
corresponding to the values of the external field that brings the states
$m$ and $m^\prime$ on-resonance;
the resonances form groups close to the fields $H_z\approx nH_1$
where $n=m+m^\prime$ and $H_1=\frac{A}{g\mu_{\rm B}}$,
see Fig.~\ref{fg:relaxation}.
The broadest of the resonances resemble those seen in usual experiments,
while most of the resonances are too narrow to be seen and, as is discussed 
below, already the phonon coupling is sufficient to suppress them.
One of the main topics of this paper is the possibility to detect some
of the satellite peaks by the application of a relatively strong
transverse field $H_\perp$ as well as to suppress the already visible 
peaks by pointing $H_\perp$ along one of the hard axes.

The whole curve in Fig.~\ref{fg:relaxation} can be understood in terms 
of the expression
\begin{eqnarray}
  \tau=\sum_n \tau_0^{(n)}D_n(H_z)e^{\beta U_n(H_z)}
\label{eq:peaked_curve}
\end{eqnarray}
where the index $n$ enumerates the resonances (pairs of states $m_n$ and 
$m_n^\prime$). As discussed in App.~\ref{app:Lorentz}, there is 
a certain $\tau_0^{(n)}$ for each resonance; the function $D_n(H_z)$ is 
a tunneling-induced Lorentzian that yields the peak shapes, and $U_n(H_z)$ 
is the effective barrier height for the $n$th resonance, i.e., 
$U_n=E_{m_n}-E_{-10}$ ($E_{-10}$ is the energy of the metastable minimum).
The widths of the Lorentzian peaks are given by
\begin{eqnarray}
  \delta H_z=\frac{4|\Delta_n|}{g\mu_{\rm B}|m_n-m_n^\prime|},
\end{eqnarray}
where $2|\Delta_n|$ is the tunnel splitting between the resonant states,
cf. App.~\ref{app:Lorentz}.
As is seen in the figure and also suggested
by the inset in Fig.~\ref{fg:energies_wis}, the background is not given by 
the over-barrier relaxation but by the ``leakage'' due to the direct coupling
of states with $m-m^\prime=\pm4$, e.g., $m=\pm2,m^\prime=\mp2$ close to 
$H_z=0$ and $m=\pm1,m^\prime=\mp3$ close to $H_z= 2H_1$. 
This effectively lowers the barrier height to $U\approx E_{-2}-E_{-10}$.
The rates that determine $\tau_0^{(n)}$, depend only weakly on temperature,
the resonant states, and $H_z$, cf. App.~\ref{app:Lorentz}. 
This is in line with the experimental observation that the estimated
prefactor of the Arrhenius law, $\tau_0$, varies within an order of magnitude
for different samples and resonances.

The sharpness of the narrowest peaks in Fig.~\ref{fg:relaxation} is 
an artefact of the reduced model used so far, i.e., neglect of the 
nondiagonal matrix elements, and let us next consider two things affecting 
these peaks: decoherence due to spin-phonon coupling and the broadening 
effect of a transversal magnetic field $H_\perp$. 
For other contributions such as the hyperfine and dipolar interactions, 
see Sec.~\ref{sec:discussion}.

\begin{figure}
\epsfxsize=8.5cm
\centerline{\epsffile{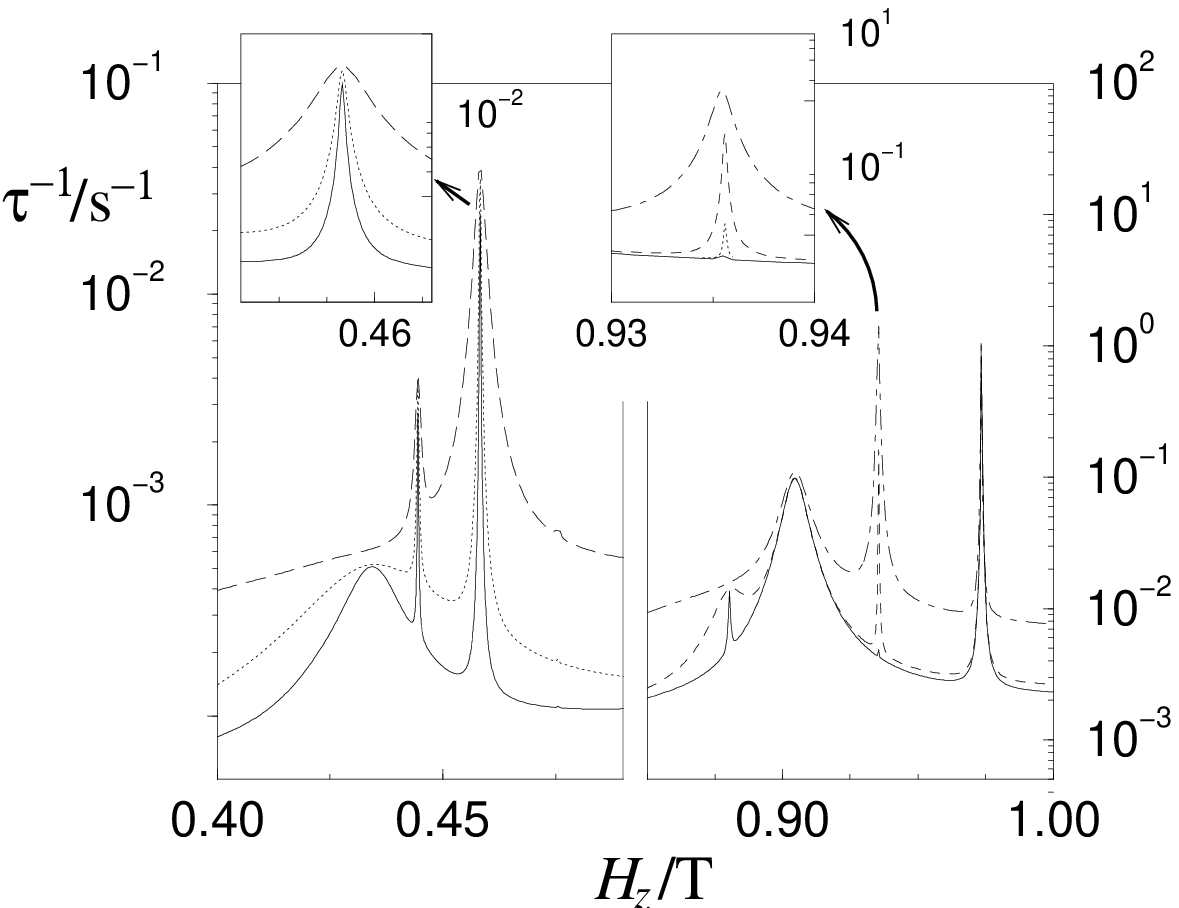}}
\caption{
  Magnification of the first and second clusters of peaks of 
  Fig.~\ref{fg:relaxation} magnified (solid line). 
  The different curves correspond to: $H_\perp=0.0$T (solid), 0.01T (dotted),
  0.05T (dashed), 0.1T (long dashed), and 0.2T (dot dashed);
  $\phi=0^\circ$ for all the curves.
  The peaks further to the right correspond to resonances
  between lower states, hence the higher maxima.
  The peaks (from left to right) correspond to the resonances 
  $m=1,..,4,m^\prime=-m-1$ in the left figure (the two first ones are merged 
  together) and $m=2,..,5,m^\prime=-m-2$ in the right figure. 
  Here we used also the offdiagonal elements of $\rho(t)$ with the result
  that some of the sharper peaks are reduced in height or even suppressed
  due to spin-phonon coupling.
  The peak widths show strong $H_\perp$-dependence except for the second and 
  fourth peaks on the right -- these arise from $B_4^2$ and $B_4^3$-type of 
  coupling.
  Once the peak widths due to the tunnel splittings exceed the phonon-induced 
  width, the peak heights are essentially determined by the energy 
  $U_n=E_n-E_{-10}$ needed to reach the resonant states.
        }
\label{fg:decohered_peaks}
\end{figure}

The decay of the nondiagonal elements of the density matrix is
a classical definition of decoherence and this is also what is 
seen here when the above calculation is performed using the full $\rho(t)$. 
Inclusion of the off-diagonal elements still produces Lorentzian
peaks but now the narrow resonances -- with widths of the same order of 
magnitude as the spin-phonon coupling -- are broadened and reduced in height. 
The narrowest peaks may even be completely suppressed,
see the solid-line curves in Fig.~\ref{fg:decohered_peaks}.
Despite its intuitive appeal, the suppression of the narrow peaks should 
be taken only qualitatively because of the limitations of the Born 
approximation used in treating the spin-phonon coupling.
For this reason, the following considerations are restricted to regimes 
where the effect of the nondiagonal elements of $\rho(t)$ is negligible.

\begin{figure}
\epsfxsize=8.5cm
\centerline{\epsffile{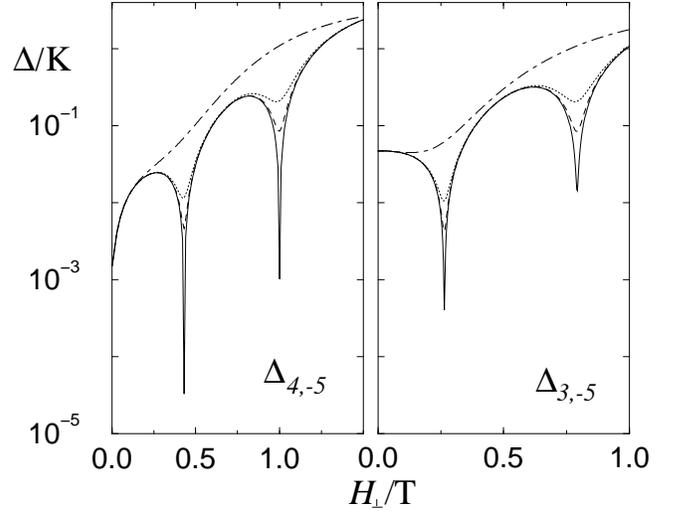}}
\caption{
  The tunnel splittings $\Delta_{4,-5}$ (left) and $\Delta_{3,-5}$ (right) 
  as a function of the transverse magnetic field when $H_z$ is kept fixed 
  at the position of the corresponding resonance,
  0.4581T and 0.9045T, respectively. 
  In both figures, the four curves correspond to
  different angles $\phi$: $0^\circ$ -- dot-dashed, $40^\circ$ -- dotted, 
  $43^\circ$ -- dasheded, and $45^\circ$ -- solid; the vertical axes are 
  the same.
        }
\label{fg:splittings23}
\end{figure}

The suppressing effect of the spin-phonon coupling being essentially 
a constant, the sharper resonances can be made observable by broadening 
them with a transverse field $H_\perp$;
even ground state tunneling has been observed in high enough fields.
\cite{Belessaetal}
The various tunnel splittings are shown in Fig.~\ref{fg:splittings} 
as a function of $H_\perp$ pointed to three different directions $\phi$.
It can be seen that for small values of the field, the splittings are 
essentially independent of the chosen angle, see also 
Fig.~\ref{fg:splittings23}. 
Furthermore, tunnel splittings between states that can be coupled with 
sole $B_4$ terms of ${\mathcal H}_{\rm T}$ are quite insensitive
to the transverse field below 0.1-0.2T. 
Figure \ref{fg:decohered_peaks} illustrates the effect of $H_\perp=H_x$ onto
the two first series of peaks increasing $H_x$ monotonously increases 
the tunnel splittings.
The resonances that initially ($H_x=0$) were broader than 
${\tau_0^{(n)}}^{-1}$ are seen to retain their height;
the narrower peaks, such as the two shown in the insets, are strongly 
broadened with increasing $H_x$ and also their heights increase once 
their widths become larger than the spin-phonon rates.

A transverse field cannot only increase the tunnel splittings but
it can also reduce them, see Fig.~\ref{fg:splittings}. 
In the ideal case, the splittings can even be suppressed by varying 
the angle $\phi$ where the transverse field is pointed.
For the zeroth peak, i.e., for $H_z\approx0T$, all ten resonances occur at the
same position and the suppression of any one of them is obscured in the
relaxation rate curves by the simultaneous broadening of other resonances.
For this reason, let us first focus on the resonances occuring at finite 
$H_z$ and return to the $H_z\approx0$ case in the next section.

Figure \ref{fg:splittings23} shows the tunnel splittings $\Delta_{4,-5}$ 
and $\Delta_{3,-5}$, respectively, as a function of $H_\perp$ and for
four different values of $\phi$. 
The former corresponds to the resonance shown in the left inset of 
Fig.~\ref{fg:decohered_peaks}, while the latter corresponds to the broad 
resonance in the right panel of Fig.~\ref{fg:decohered_peaks} which is also
the one seen in experiments at $H_z\approx 2H_1$.
Figure \ref{fg:splittings23} also exemplifies some general properties of
the tunnel splittings. As already mentioned above, the $\Delta$'s depend 
only weakly on $\phi$ for low $H_\perp$, while for larger fields and 
$\phi$ close to the hard axes, the tunneling amplitudes are successively 
increased and reduced as $H_\perp$ is increased.
On the other hand, with $H_\perp$ held fixed to one of the suppression points,
the $\Delta$'s oscillate as a function of $\phi$.
Similar to recent work on Fe$_8$, see e.g. 
Ref.~\onlinecite{Wernsdorfer_Science}, 
these phenomena are attributed to alternating constructive and destructive 
interference of the geometrical phase in the tunneling amplitudes.
\cite{Lossetal,Garg}
The number of minima for a given $\Delta_{m,m^\prime}$ appears to be, 
as was also pointed out in Ref.~\onlinecite{review}, given by the number 
of $B_4$-terms involved in coupling the states $m$ and $m^\prime$. This ranges
from zero at the top of the barrier to five for the gound states $m=\pm10$, 
cf. Fig.~\ref{fg:splittings}.

\begin{figure}
\epsfxsize=8.5cm
\centerline{\epsffile{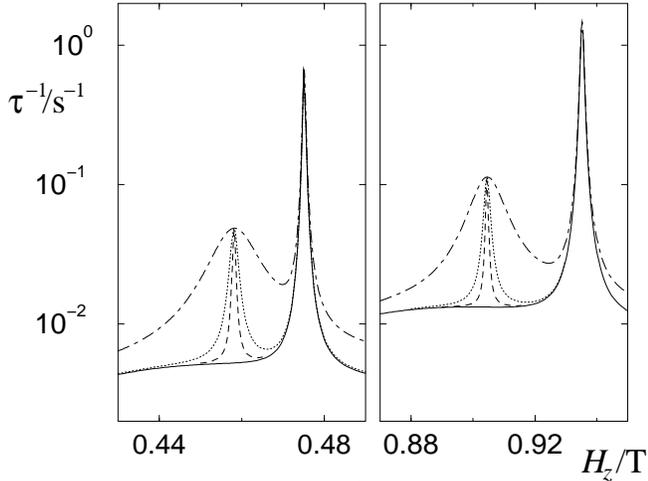}}
\caption{
  Relaxation rate $\tau^{-1}$ as in Fig.~\ref{fg:decohered_peaks} 
  but now for stronger transverse field $H_\perp$ and for different 
  angles $\phi$. 
  The line types and angles are as in Fig.~\ref{fg:splittings23}
  and the values of $H_\perp=0.4311$T and 0.2643T are chosen to match 
  the first suppression points (for the actual suppression occurring 
  at the minimum value for $\Delta$ the values for $H_\perp$ should be 
  given with even higher accuracy). Temperature is 2.5K.
        }
\label{fg:suppressions23}
\end{figure}

Figure \ref{fg:suppressions23} shows the relaxation rates corresponding 
to the resonances of Fig.~\ref{fg:splittings23} with $H_\perp$'s
chosen to match the first points of suppression $H_\perp^{m,m^\prime}$.
The resonance width is found to be very sensitive to $\phi$ quite as 
expected and, for $\phi=45^\circ$, the left-most peak becomes suppressed.
Note, that the suppression is also quite sensitive to the value of $H_\perp$
as well and the higher peaks in Fig.~\ref{fg:suppressions23} are hardly
affected at all by changes in $\phi$.

\subsection{Ac-susceptibility}
\label{sec:results_chi}

\begin{figure}
\epsfxsize=8.5cm
\centerline{\epsffile{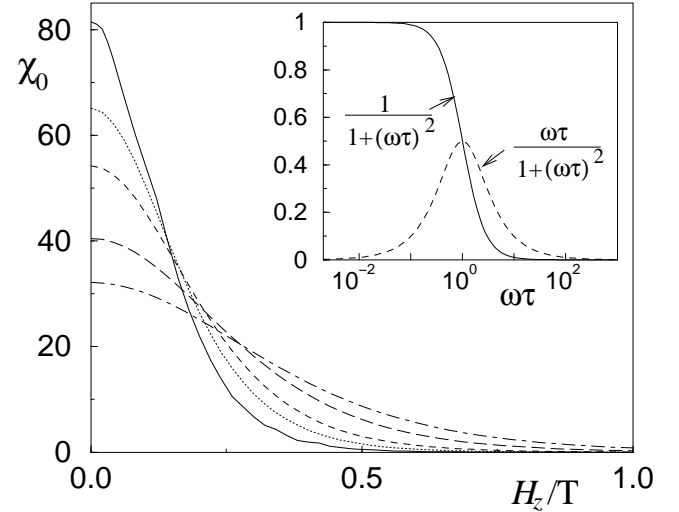}}
\caption{
  Static susceptibility $\chi_0$ as a function of the longitudinal magnetic
  field shown for different temperatures: 2K (solid), 
  2.5K (dotted), 3K (dashed), 4K (long dashed), and 5K (dot-dashed).
  The dynamic susceptibility $\chi(\omega)$ is essentially composed of 
  $\chi_0$ multiplied by the real and imaginary parts of 
  $1/(1-i\omega\tau)$ shown in the inset as the solid and dashed
  lines, respectively.
        }
\label{fg:chi_comp}
\end{figure}

In explaining experimental results, the susceptibility $\chi(\omega)$ 
is often described by the formula
\begin{eqnarray}
 &&\chi(H,T,\phi,\theta,\omega)=
  \frac{\chi_0(H,T,\theta)}{1-i\omega\tau(H,T,\phi,\theta)}
 \nonumber\\[-5pt]
 &&
\label{eq:chi_experimental}
\end{eqnarray}
where $\tau(H,T,\phi,\theta)$ is the relaxation time, 
$\omega$ the frequency of the small excitation field, and
$\chi_0(H,T,\theta)$ the static susceptibility, 
cf. Ref.~\onlinecite{Luisetal1}.
We emphasize here the $\phi$-dependence as we wish to investigate 
the interference effects and resulting oscillations in $\tau_1$
(below we do not explicitly write the parametric dependences of $\tau$ but
they are implicitly assumed).
Equation (\ref{eq:chi_experimental}) relates to the results of 
Sec.~\ref{sec:chi} as follows. Equation (\ref{eq:chi_exp}) formally 
accounts for all the modes $i$ and for each mode divides into two 
contributions, one being $1/(1-i\omega\tau_i)$ and 
the other the respective prefactor.
It turns out that, for most parameter values, the over-barrier relaxation, 
i.e., the mode $i=1$ strongly dominates over the others and 
one can use Eq.~(\ref{eq:chi_experimental}) to a good approximation.
Figure \ref{fg:chi_comp} illustrates the two contributions to 
Eq.~(\ref{eq:chi_experimental}); 
they form the basis for understanding the following results.
In all figures, the susceptibility is expressed for a single spin 
and in units of ${\rm KT}^{-2}$.

In the present model, the main correction to Eq.~(\ref{eq:chi_experimental}) 
corresponds to an intra-valley mode describing transitions between 
the two lowest states on the same side of the anisotropy barrier. 
This correction increases with increasing $H_z$ and temperature and,
for the parameters and the curves of $\chi_0$ in Fig.~\ref{fg:chi_comp}, 
becomes of the same order of magnitude as the actual relaxing mode 
once $H_z\approx0.9-1$T.
However, such a mode corresponds to very high frequencies and is
neglected below. 
There could be also other sources of high-frequency contributions 
such as dipole-dipole flip-flops, nuclear spin dynamics, or 
moving impurities but these are beyond the scope of this work.

\begin{figure}
\epsfxsize=8.5cm
\centerline{\epsffile{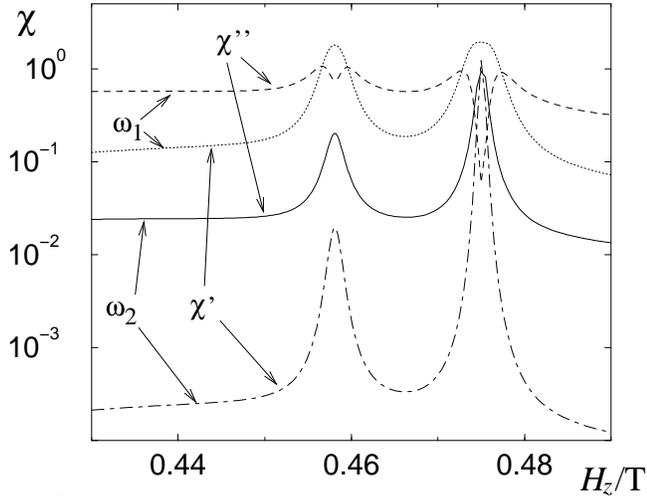}}
\caption{
  The real and imaginary parts of susceptibility for the same parameters
  as the dotted curves in the left panels of Fig.~\ref{fg:suppressions23}, 
  i.e., $H_\perp=0.4311$T, $\phi=40^\circ$, and $k_{\rm B}T=2.5$K.
  The curves correspond to frequencies $\omega_1=0.02$Hz -- 
  $\chi^\prime$ (dotted), $\chi^{\prime\prime}$ (dashed) --
  and $\omega_2=0.5$Hz -- $\chi^\prime$ (dot-dashed), $\chi^{\prime\prime}$ 
  (solid). The frequencies were chosen such that $\omega_1<\tau^{-1}$ for 
  the both of the peaks in Fig.~\ref{fg:suppressions23} -- hence the notches
  in $\chi^{\prime\prime}$, while $\omega_2>\tau^{-1}$ for all $H_z$.
  }
\label{fg:chi_example}
\end{figure}

As to the actual results, it is in principle sufficient to combine 
the curves for $\tau$ of the previous section with those shown in 
Fig.~\ref{fg:chi_comp}.
The shapes of the real and imaginary parts of $1/(1-i\omega\tau)$
readily imply how the resulting susceptibility behaves if one fixes 
$\omega$ --
it turns out that all the structure found in $\tau^{-1}$ in the previous
section can also be found in $\chi(\omega)$.
First, the response is the most sensitive to changes in $\tau$ when
$\omega\approx\tau^{-1}$ or $\omega\tau\approx1$, cf. the inset of 
Fig.~\ref{fg:chi_comp}.
Second, ${\rm Re}(\chi(H_z;\omega))$ replicates the shape of $\tau(H_z)$ and
exhibits peaks and valleys as $\tau^{-1}$ increases and decreases.
This is the case with the imaginary part, ${\rm Im}(\chi(H_z;\omega))$,  
as well, but only as long as $\omega\tau\ge1$, i.e., as long as 
we remain on the right hand side of the peak in ${\rm Im}(\chi(H_z;\omega))$.
For lower $\omega$,  $\omega\tau$ can cross the maximum point and the picture
with the peaks and valleys turns upside down.
These points are illustrated in Fig.~\ref{fg:chi_example} for a case
corresponding to one of the curves in \ref{fg:suppressions23}.

\begin{figure}
\epsfxsize=8.5cm
\centerline{\epsffile{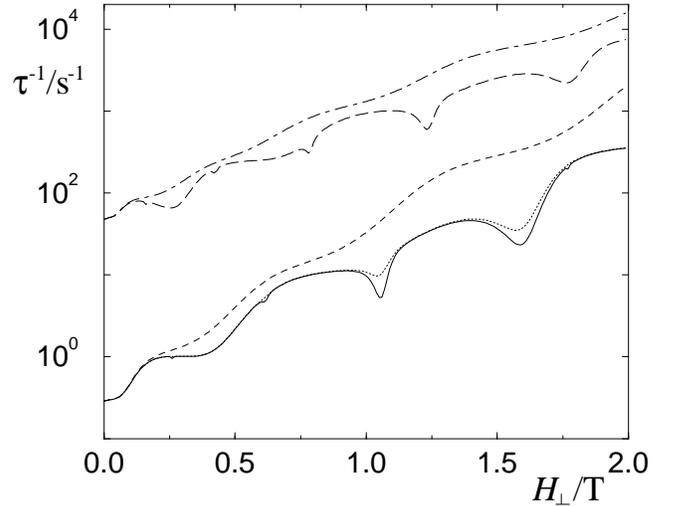}}
\caption{
  Oscillations in the relaxation rate as a function of $H_\perp$.
  The lower and upper groups of curves correspond to 
  $H_z=0.0002{\rm T},k_{\rm B}T=3.0{\rm K}$ and
  $H_z=0.01{\rm T},k_{\rm B}T=5.0{\rm K}$, respectively.
  The individual curves correspond to different angles $\phi$:
  $0^\circ$ (dashed and dot-dashed), $43^\circ$ (dotted), and
  $45^\circ$ (solid and long dashed).
        }
\label{fg:tau_oscs}
\end{figure}

Let us next consider a different scheme, keeping $H_z$ and $\phi$ fixed and
varying $H_\perp$ starting from zero field.
The case $H_z\approx0$T is of particular interest because it allows comparison
to recent experiments by Wernsdorfer done on the related material Fe$_8$,
cf. Ref.~\onlinecite{Wernsdorfer_LT} -- the experimental data showed clear
oscillations of the relaxation rates as a function of $H_\perp$ in 
the thermally activated regime.
By choosing parameters in the feasible range of these experiments,
we find somewhat similar oscillations also for Mn$_{12}$.
Figure \ref{fg:tau_oscs} shows first the relaxation rates for two different
combinations of temperature and the longitudinal field: $T=3.0$K and 
$H_z=0.2$mT to show the behaviour close to (or at) the peak maximum and 
at a lower temperature, and $T=5.0$K and $H_z=10$mT as an example of 
the behaviour further away from the maximum and at a higher temperature.
Also the $\phi$ dependence is shown.
We would like to stress the novelty of this result:
such oscillations have neither been observed nor predicted before.

\begin{figure}
\epsfxsize=8.5cm
\centerline{\epsffile{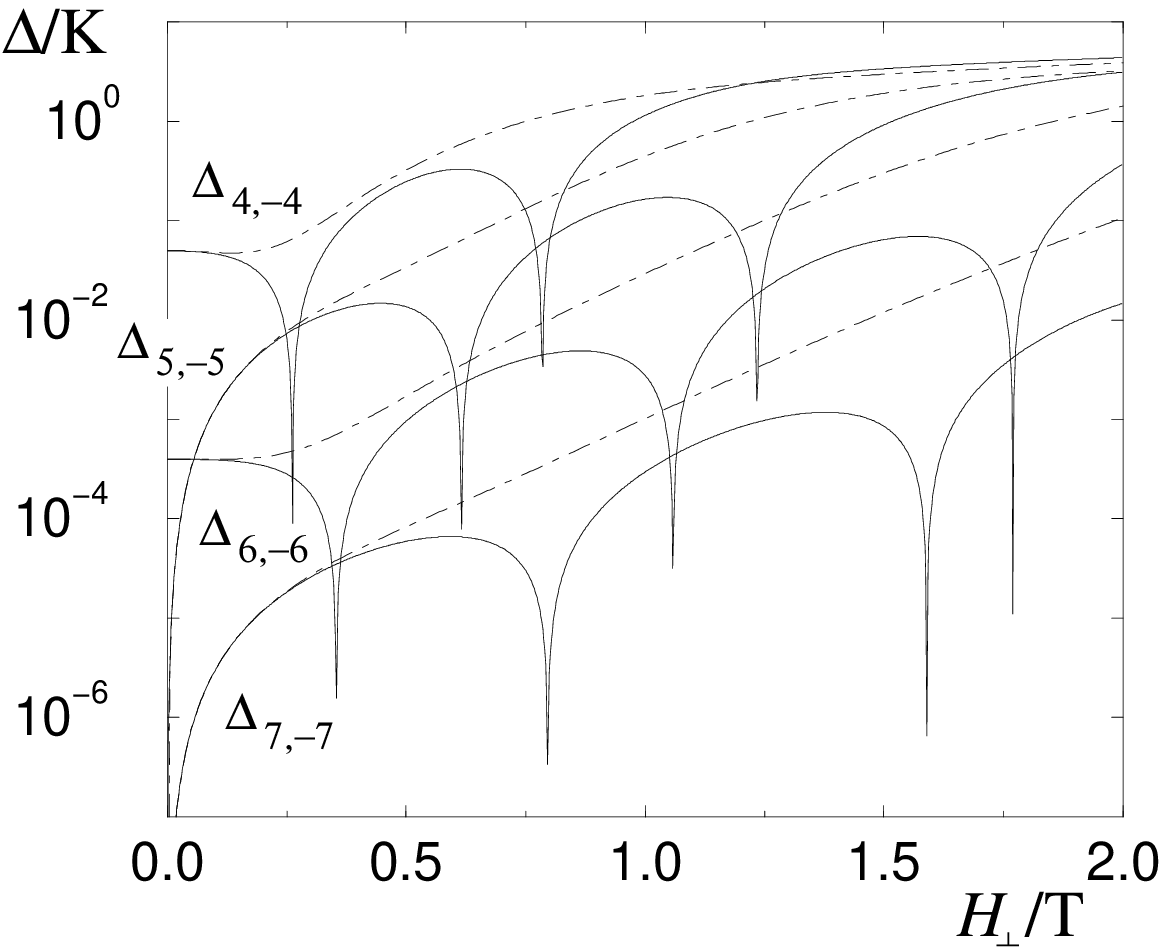}}
\caption{
  Tunnel splittings $\Delta_{m,-m}$ for four resonances relevant
  for Fig.~\ref{fg:tau_oscs}. The dot-dashed and solid lines denote
  the angles $\phi=0^\circ$ and $45^\circ$, respectively.
        }
\label{fg:split_oscs}
\end{figure}

\begin{figure}
\epsfxsize=8.5cm
\centerline{\epsffile{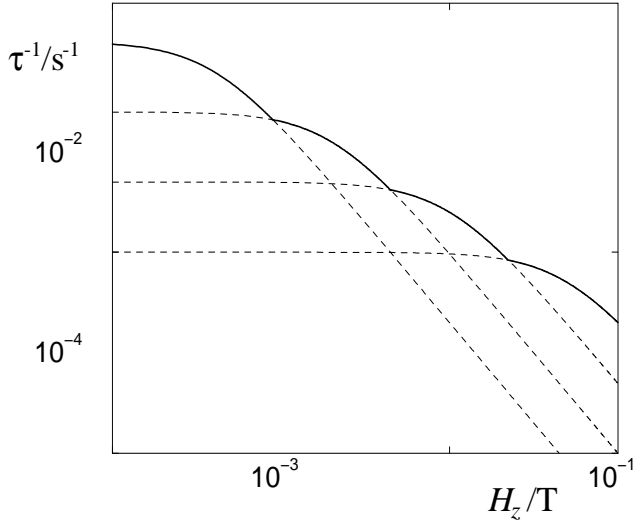}}
\caption{
  Log-log scale schematic of Lorentzian curves on top of each other 
  similar to the $H_z\approx 0$T situation for Mn$_{12}$.
  The higher the curves are the narrower they get and, on the other hand,
  the higher the $H_z$ the lower the observed resonances are.  
  The relaxation rate is determined by the fastest possible rate,
  depicted in the figure with the thick solid line.  
        }
\label{fg:stacked_Lorentz}
\end{figure}

For $\phi=0^\circ$, the gentle oscillations resembling a strongly-smeared 
staircase stem from the fact that the lower (in terms of energy; higher
in terms of rates) resonances are broadened and one by one start to dominate 
the relaxation, see the dot-dashed curves in Fig.~\ref{fg:split_oscs} for 
the tunnel splittings and Fig.~\ref{fg:stacked_Lorentz} for the general idea.
All the additional structure, seen for $\phi=45^\circ$, is due to 
the oscillations in the tunnel splittings and the suppression of some 
of the resonances.
The positions of the notches can be compared with the structure 
of $\Delta_{m,-m}$ in Fig.~\ref{fg:split_oscs} and one finds that 
for smaller $H_z$ the relaxation takes place via the lower-lying 
resonances such as $m=\pm6$ and $m=\pm7$; further away from the maximum,
the broader resonances between states $m=\pm4$ and $m=\pm5$ act as 
the dominant relaxation paths.

\begin{figure}
\epsfxsize=8.5cm
\centerline{\epsffile{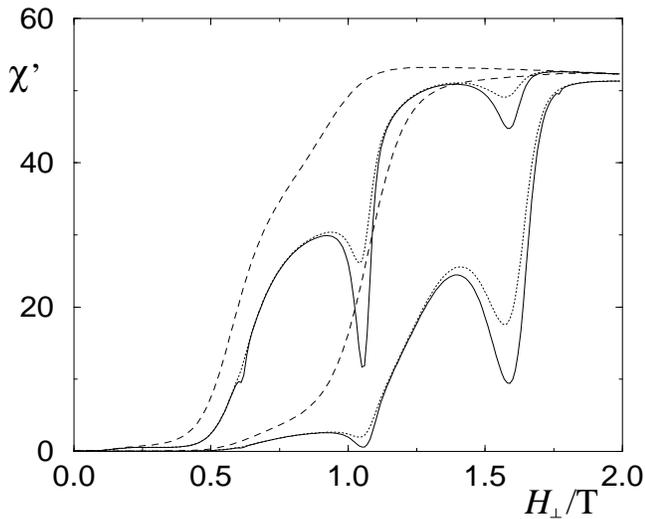}}
\caption{
  Real part of the susceptibility for $H_z=0.2$mT, $k_{\rm B}T=3.0$K, and
  for the angles $\phi=0^\circ$ (dashed), $43^\circ$ (dotted),
  and $45^\circ$ (solid). The upper three curves correspond to
  the frequency $\omega=10$Hz and the lower ones to $\omega=50$Hz.
  The line types correspond to those for $\tau^{-1}$ in 
  Fig.~\ref{fg:tau_oscs}.
        }
\label{fg:oscs2a}
\end{figure}

\begin{figure}
\epsfxsize=8.5cm
\centerline{\epsffile{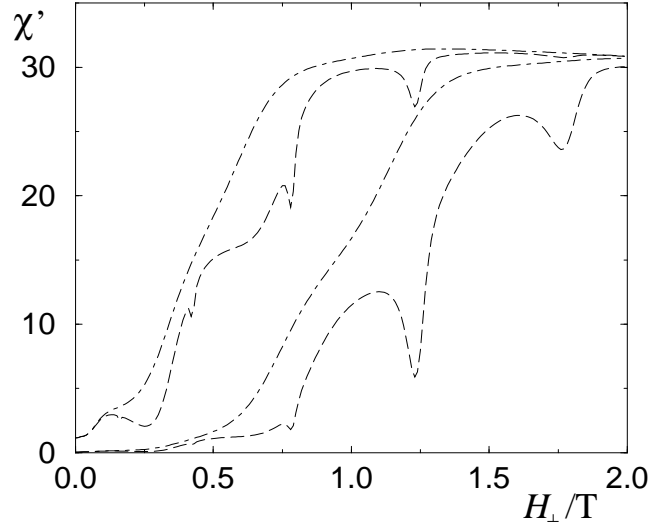}}
\caption{
  Real part of the susceptibility for $H_z=10$mT, $k_{\rm B}T=5.0$K, and
  for the angles $\phi=0^\circ$ (dot-dashed) and $45^\circ$ (long dashed),
  i.e., the line types correspond to those in Fig.~\ref{fg:tau_oscs}. 
  The upper pair of curves corresponds to
  $\omega=250$Hz and the lower one to $\omega=1250$Hz.
        }
\label{fg:oscs2b}
\end{figure}

Figures \ref{fg:oscs2a} and \ref{fg:oscs2b} show the real part of 
the susceptibility $\chi^\prime$ corresponding to the two cases 
of Fig.~\ref{fg:tau_oscs}. 
The purpose of the different frequencies is to show that
also here one can choose the structure of interest and study it 
by tuning the frequency to fulfill $\omega\tau\approx1$.
In both figures, there are regimes where the susceptibility can 
be varied by a factor of five; in Ref.~\onlinecite{Wernsdorfer_LT}
the oscillations were quite clear already with the amplitude being 
a mere 20\% of the signal.

\subsection{Discussion -- relevance to experiments}
\label{sec:discussion}

So far we have considered the simple model comprising a single spin
coupled to a phonon bath but, as was pointed out in the introduction, 
in real samples there are also other kinds of interactions.
In this section, we consider the additional features arising from 
the hyperfine and/or dipolar interactions and aim to point out 
the experimentally relevant aspects of the results obtained above.

Let us first recall some experimental facts concerning relaxation
measurements and results -- these underlie also the understanding 
and appreciation of the susceptibility measurements. 
Relaxation rates are typically measured by first magnetizing the sample to
saturation, and then reversing the direction of the field 
and measuring the resulting magnetization as a function of time.
The initial relaxation is observed to be nonexponential -- this is attributed 
to dipolar interactions, see below -- while, at later times, it becomes 
exponential. 
\cite{Thomasetal2}
Several authors have proposed an extended exponential 
$M(t)=M(0)\exp[-(t/\tau)^\beta]$ to account for both of these regimes
with just one additional fitting parameter $\beta$. 
In Ref.~\onlinecite{Thomasetal2} it was found that $\beta$ varies
from $\beta\approx0.5$ below 2.0K to $\beta\approx1$ -- usual exponential
relaxation -- roughly above 2.4K.  
The thus obtained relaxation rates exhibit a series of broad
Lorentzian-shaped resonances; their height and location correspond to 
tunneling-assisted relaxation 3-4 levels below the top of the barrier.

This shows two clear differences as compared to the present work:
in experiment, the relaxation may be nonexponential even though 
the single-spin model always yields exponential behaviour, and no satellite 
peaks are observed (see Ref.~\onlinecite{Sarachik-Zhong} for exceptions). 
In order to understand these discepancies, let us first consider the effect 
of the nuclei via the hyperfine interactions and then the intermolecular 
dipolar interactions.

{\bf Hyperfine interactions.}
In Mn$_{12}$, all the manganese nuclei have magnetic momenta and
the hyperfine interaction between the nuclei and the molecular
spin state is relatively large, of the order of 10mT.
Recently, several authors have investigated how this affects
tunneling and the relaxation in Mn$_{12}$.
\cite{Hartmanetal,Garanin-Chudnovsky,Prokofev-Stamp,Garaninetal,review}
For the present purposes, the relevant effect of the hyperfine interactions 
is to induce an {\it intrinsic} Gaussian broadening, of the width 
$\sigma_{\rm hyp}\approx6$mT,
\cite{Wernsdorfer_EPL} 
to all levels, i.e., the nuclear spins are importantly dynamic
and their influence on the molecular spin
cannot be reduced to a rigid but spatially varying background field.
Simultaneously with the broadening, the hyperfine interactions
reduce the tunneling amplitudes of the resonances for which 
$\Delta<\sigma_{\rm hyp}$ -- this should lead to reduced peak heights 
in $\tau^{-1}(H_z)$.

With this in mind, let us consider the different regimes in terms of 
the relative magnitudes of the tunnel splitting and the hyperfine broadening.
First, for resonances with $\Delta\gg\sigma_{\rm hyp}$ the shapes of 
the resonances are expected to be Lorentzian with the widths determined 
by the tunnel splittings $\Delta$. 
This is the regime, where all our results apply.
In the other extreme, $\Delta\ll\sigma_{\rm hyp}$, the resonances should 
be essentially suppressed providing one possible explanation 
why the sharp satellite peaks are not observed in experiments.
Note that the minuscule phonon-induced broadening or dephasing
is hidden under the hyperfine broadening and cannot be seen. 
In this regime, one could in principle try and extend the present theory by
adding by hand a strong dephasing term to the nondiagonal density matrix 
elements.
The intermediate regime where $\Delta\approx\sigma_{\rm hyp}$
is the most interesting of the three cases.
In this regime, the peak shape should be a combination of Lorentzian and 
Gaussian curves and, depending on which one of $\Delta$ or 
$\sigma_{\rm hyp}$ is larger, one of the shapes should dominate.
In the tail region, i.e., away from the peak maxima, the Lorentzian tails 
dominate and it has been suggested that this together with experimental 
error bars may obscure the resolution between the two types of curves, 
cf. e.g. Ref.~\onlinecite{review}.

The immediate conclusion from these considerations is that, 
if the application of $H_\perp$ broadens some of the tunnel splittings
$\Delta_{m,m^\prime}$ to exceed $\sigma_{\rm hyp}$, 
the corresponding resonance should become observable.
If, on the other hand, the transverse field is applied along one of 
the hard axes, a given resonance becomes suppressed for certain special
values of $H_\perp$; in the presence of the hyperfine interactions
this should happen already when $\Delta_{m,m^\prime}$ becomes smaller 
than $\sigma_{\rm hyp}$.
The intermediate regime can be intentionally achieved by tuning the tunnel 
splitting from being well below $\sigma_{\rm hyp}$ to above it.
This may provide means to probe the Gaussian broadening,
see also the subsection below focusing on the advantages of 
susceptibility measurements.

{\bf Dipolar interactions.}
The intermolecular spin-spin interactions are of dipolar form
and they are weaker in Mn$_{12}$ than, e.g., in Fe$_8$.
Due to their short range, the dipolar fields can vary in space 
changing the local field at the position of the individual molecules.
In our view, the essential difference between the hyperfine and dipolar 
interactions can be stated as follows: 
even if one could measure the response of a single molecule,
this would always be dressed by the level broadening and reduction 
in tunneling amplitudes due to hyperfine interactions intrinsic to each 
molecule; 
the dipolar fields, on the other hand, just change the molecule's local 
electromagnetic environment.

In experiment, the relaxation of the magnetization $M(t)$ leads to 
time-dependent dipolar fields and, in order to describe the relaxation 
correctly, it would be necessary to solve for $M(t)$ self-consistantly,
for simulations see e.g. Refs.~\onlinecite{Prokofev-Stamp} and 
\onlinecite{Ohmetal}.
However, it is this time-dependent field that provides 
an explanation to the initial nonexponential relaxation.
For example in Ref.~\onlinecite{Thomasetal2}, it is therefore concluded 
that the deviation from a single-exponential relaxation, i.e., $\beta\neq1$,
demonstrates the important role of the dipolar interactions and dynamics of 
the spin distribution.
It should be kept in mind, though, that also a static distribution of 
local fields (be it dipolar or not) 
-- and hence relaxation rates $\tau^{-1}(H_{z}^{\rm local})$ -- 
leads to a superposition of exponential rates, which looks nonexponential.
Such a distribution of fields also hides all features in
$\tau^{-1}(H_z)$ that are sharper than this distribution.

The time-dependence of the dipolar fields owes to the fact that 
the sample is first magnetized and, as the field direction is abruptly
reversed, the dipolar distribution finds itself far from equilibrium 
and quickly starts to relax.
The reason for such experiments is the strong response from almost 
all the spins.

In anticipation of the discussion on susceptibility, let us consider
the dipolar distributions at equilibrium. By distribution we mean spatial 
variations in the dipolar field at the locations of the individual molecules.
First, for $H_z\approx0$T, 
the annealed (not quenched) distribution is random but due to the low 
temperatures, $k_{\rm B}T\ll E_{\pm9}-E_{\pm10}$, almost all the spins 
are aligned with the easy $z$-axis -- randomly pointing to the positive 
and negative directions -- and only contribute to the local longitudinal field.
On the other hand, the equilibrium magnetization is close to saturation
already for $H_z\approx H_1$, thus drastically narrowing the dipolar 
distribution -- e.g. for $T=3.0$K (5.0K) and $H_z=H_1$, more than 95\% 
(85\%) of all the spins are aligned parallel to $H_z$ and
all the molecules feel essentially the same field.
Such distributions have been experimentally verified in Fe$_8$,
cf. Ref.~\onlinecite{Wernsdorfer_Fe8}.

{\bf Susceptibility.}
The influence of the dipolar dynamics on relaxation can be avoided 
almost completely by measuring linear response to a small ac-field, i.e., 
the ac-susceptibility, instead of $M(t)$. 
This has the advantage that the system is probed in its equilibrium 
state and ideally by a small enough field that in itself does not 
perturb the equilibrium.
Therefore we propose that the susceptibility measurements provide
a gentle or noninvasive means to probe the relaxation dynamics in 
absence of the time-dependent dipolar distribution.
Furthermore, while the hyperfine interactions cannot be tuned, 
the static distribution can be made markedly narrower by a finite $H_z$,
see above. 
As on the other hand $\chi_0$ decreases with increasing $H_z$,
the first group of resonances, close to $H_1$, is especially attractive 
for investigating the oscillations in the relaxation
rates as well as the hyperfine fields themselves.
All of the resonances around $H_1$ depend strongly on $H_\perp$ and 
can be broadened such that $\Delta>\sigma_{\rm hyp}$ making them
observable; the peaks can also be selectively suppressed if 
$\phi\approx45^\circ$. 
The hyperfine fields may even simplify the observation of the suppressions 
as very narrow peaks are strongly reduced in height.
For a sharp dipolar distribution, we expect that also the crossover 
between Lorentzian and Gaussian shapes of $\tau^{-1}(H_z)$ 
should be observable when changing the tunnel splittings with 
the transverse field.

\section{Conclusions}
\label{sec:conclusions}

To conclude, we present a diagrammatic description of the spin dynamics
of the molecular magnet Mn$_{12}$.
The work focuses on the regime of thermally-activated tunneling, i.e.,
$T>2.0{\rm K}$, and emphasizes the phenomena that could be observed
for strong transverse magnetic fields.
In the calculations, we study the dynamics of a single spin $S=10$ 
coupled to a phonon bath.
The role of the phonons is in the thermal activation of the spins 
to states with higher energies and larger tunneling amplitudes.

As the first main result, we calculate the dynamic susceptibility 
$\chi(\omega)$ starting from the same microscopic Hamiltonian as is used 
for the relaxation rates. 
Susceptibility is found to reflect the rich structure found in 
$\tau^{-1}(H_z)$ and we argue that susceptibility measurements 
are in fact more sensitive and better controlled in terms of time scales
and also the dipolar interactions than the relaxation experiments.

All the results obtained are calculated using the eigenbasis of the spin 
Hamiltonian, which naturally accounts for strong transverse magnetic fields.
A strong transverse magnetic field enhances tunneling through the anisotropy 
barrier and enables relaxation via eigenstates further away from the top of 
the barrier.
In relaxation or susceptibility measurements, this would lead to 
shifted and higher resonances.
The tunnel splittings are found to be very sensitive to the azimuth angle 
$\phi$ of the transverse field $H_\perp$.
It is found that, in the directions $\phi=\pi(2n+1)/4$,
the tunnel splittings exhibit alternating minima and maxima and become 
totally suppressed at certain values of $H_\perp$.
This phenomenon attributed to the interference of the geometrical or Berry
phase of alternative tunneling paths, with a destructive interference 
leading to the suppressions.
As the second major result, we predict that these oscillations in 
the tunnel splittings should be observable both in the relaxation rates 
and the susceptibility.

\section*{Acknowledgements}

We are grateful to Myriam Sarachik and Yicheng Zhong as well as Wolfgang 
Wernsdorfer for the discussions on their experiments and
M.S. and Y.Z. for providing us with their unpublished data.
We would also like to thank Michael Leuenberger and Daniel Loss
for the fruitful exchange of theoretical ideas. 
The calculations were in part carried out in
the Center for Scientific Computing (CSC) in Finland.
This work has been supported by the Finnish Academy of Science and Letters,
the Finnish Cultural Foundation, the EU TMR network ``Dynamics of 
Nanostructures'', the Swiss National Foundation, and DFG through SFB 195.

\appendix

\section{Diagrammatic rules}
\label{app:rules}

In this Appendix, we list the rules for evaluating the diagrammatic 
expressions for the kernels $\Sigma=\int dt\Sigma(t)$ arising in 
the evaluation of 
$\langle S_z(t)\rangle$ and the kinetic equation. We assume that 
the Markov approximation is valid and check the results for 
self-consistency.

\subsection{Real-time representation}

\begin{figure}
\epsfxsize=4.5cm
\centerline{\epsffile{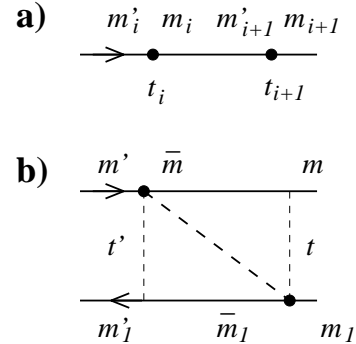}}
\caption{
  Sample diagram for illustrating the rules above. The vertical dashed lines
  in (b) are guides for the eye: the rules in the text are used to evaluate
  the piece of the diagram between times $t^\prime$ and $t$ that
  contributes to $\Sigma(t^\prime-t)$.
       }
\label{fg:rules}
\end{figure}

In the time domain, the rules for $\Sigma(t)$ are
\begin{enumerate}
\item
  {\bf Propagators}\\
  Assign a factor $\exp[-i{\mathcal H}_S(t_{i+1}-t_i)]$ to each piece of 
  a forward propagator between two vertices at times $t_i$ and $t_{i+1}$ 
  $(t_i<t_{i+1})$. 
  For a particular diagram with specific states at the ends of 
  the propagators, cf. Fig.~\ref{fg:rules}a, take the element 
  \begin{eqnarray}
    \left[e^{-i{\mathcal H}_S(t_{i+1}-t_i)}\right]_{m_{i+1}^\prime,m_i}.
  \label{eq:propagator_m}       
  \end{eqnarray} 
  For a backward propagator, change the sign in the exponent and 
  invert the order of the states $m_i$ and $m_i^\prime$.
  In the $m$-basis, the tunneling contained in ${\mathcal H}_S$ changes 
  the states along the propagator but in the $d$-basis the above exponential 
  reads
  \begin{eqnarray}
    e^{-iE_{d_i}(t_{i+1}-t_i)}.
  \label{eq:propagator_d}
  \end{eqnarray} 
  A compact way to account for the tunneling in Eq.~(\ref{eq:propagator_m}) 
  is to rewrite it with the help of Eq.~(\ref{eq:propagator_d}) as
  \begin{eqnarray}
    \sum_{d_i} \langle m_{i+1}^\prime|d_i\rangle 
        e^{-iE_{d_i}(t_{i+1}-t_i)}
        \langle d_i|m_i\rangle.
  \label{eq:propagator_md} 
  \end{eqnarray} 
  In so doing, we come to incorporate the tunneling elements exactly
  to the corresponding diagram.
\item
  {\bf Vertices}\\
  Assign a prefactor $-i$ (+i) to each vertex lying on the forward 
  (backward) propagator.
  For each pair of vertices at times $t_i$ and $t_{i+1}$ and coupled
  by a phonon line, cf. Fig.~\ref{fg:rules}b, assign a factor 
  \begin{eqnarray}
    G_{\bar{m} m^\prime,m_1 \bar{m}_1}=
        \sqrt{S_{\bar{m},m^\prime} S_{m_1,\bar{m}_1}}\cdot 
        C_{\xi,\xi^\prime},
  \end{eqnarray}
  where $\xi=\bar{m}-m^\prime$ and $\xi=m_1-\bar{m}_1$,	
  see App.~\ref{app:sigma}.
\item
  {\bf Spin-phonon interaction lines}\\
  A phonon line connecting a pair of vertices corresponds to a phonon
  correlation function -- this is worked out in detail in 
  App.~\ref{app:sigma}. It is more convenient to first calculate this in 
  the energy representation
  \begin{eqnarray}
    \Gamma(\omega)
        = \frac{A^2}{12\rho c^5\hbar^4}\cdot\frac{\omega^3}
                {e^{\beta\omega}-1}
    \label{eq:correlator_E_app}
  \end{eqnarray}
  and then Fourier transform it
  \begin{eqnarray}
    \Gamma(t-t^\prime)&&=
    \int_{-\infty}^\infty d\omega\;\Gamma(\omega)\;e^{i\omega (t-t^\prime)}\\
    \equiv&&\int_{-\infty}^\infty d\omega 
            \frac{A^2}{12\rho c^5\hbar^4}\cdot
	  \frac{\omega^3}{e^{\beta\omega}-1}\cdot e^{i\omega (t-t^\prime)}.
  \label{eq:Fourier_Gamma}
  \end{eqnarray}
  Here $A$ is the same as in ${\mathcal H}_z$, cf. Ref.~\onlinecite{LL},
  $\rho=1.83\cdot 10^3{\rm kg/m}^3$ is the density of Mn$_{12}$,
  $c$ is the sound velocity (the only ``fitting'' parameter of the theory),
  and $\omega$ is the phonon energy, $\omega>0$ meaning absorption and
  $\omega<0$ emission.
  Note that the prefactor here differs from that found in the literature
  for the spin-phonon rates. This is just because for convenience we assign 
  part of it to the vertices, cf. Eqs.~(\ref{eq:Gs})-(\ref{eq:Cs}).
  Fourier transforming the correlator into the time domain gives
  \begin{eqnarray}
    \label{eq:correlator_t}
    && \Gamma(\sigma t)
        = \frac{A^2}{12\rho c^5\hbar^4}\cdot\\
    &&\;\;\;\;\;\;
        \cdot\Big\{\!\!-\!2\left(\frac{\pi}{\beta}\right)^4\!
        \left[\frac{1+2\;{\rm ch}^2\left(\frac{\pi\sigma t}{\beta}\right)}
                   {{\rm sh}^4\left(\frac{\pi\sigma t}{\beta}\right)}\right]
        +i\;\pi\;\delta^{\prime\prime\prime}(\sigma t)\Big\}\nonumber
  \end{eqnarray}
  where $\sigma=+1$ (-1) if the vertex with the earlier time lies
  on the forward (backward) propagator;
  the function $\delta^{\prime\prime\prime}(t)$ is the third time derivative
  of the delta function. This expression diverges at $t=0$ but this is just 
  an artefact that is removed by including a high-energy cutoff in
  Eq.~(\ref{eq:Fourier_Gamma}). For instance, an exponential cutoff
  $e^{\omega/D}$ would lead to the replacement $t\rightarrow t-i/D$ in 
  Eq.~(\ref{eq:correlator_t}) and thus avoiding the divergence.
  We are not interested in the limit $t\rightarrow 0$ 
  but rather the longer-time behaviour of the correlations:
  the correlator decays exponentially on the time scale of $1/k_{\rm B}T$
  which is 5-10 orders of magnitude faster than typical relaxation times
  $\tau$ of the reduced density matrix.
  This observation is used to justify the Markov approximation in the text.
\item
  {\bf Summations and integrations}\\
  Sum over the states internal to the diagram. For example, in evaluating
  the element $\Sigma_{mm_1,m^\prime m_1^\prime}$ in Fig.~\ref{fg:rules}b, 
  sum over the states $\bar{m}$ and $\bar{m}_1$, and integrate over all 
  the times (time differences) in the diagram.
\end{enumerate}  

We might add a fifth rule to capture a general property of combinations of
diagrams: a series of irreducible diagrams is evaluated iteratively such 
that all the information needed from earlier processes/diagrams is 
incorporated into the reduced density matrix $\rho(t)$.
As an example, the diagram in Fig.~\ref{fg:rules}b equals
\begin{eqnarray}
\label{eq:diagram_t}
 && -i\cdot i \sum_{\bar{m},\bar{m}_1}
    G_{\bar{m} m^\prime,m_1 \bar{m}_1}\\
 && \;\;\;\;\;\;\;\;\;\;\;\;\;\;\;\;\;\;
        \cdot\sum_{d,d_1}
        \langle m_1^\prime|d_1\rangle
        \langle d_1|m_1\rangle
        \langle m|d\rangle
        \langle d|\bar{m}\rangle\nonumber\\
 && \;\;\;\;
        \cdot\int_0^t dt^\prime e^{-i(E_d-E_{d_1})(t-t^\prime)}
        \Gamma(t^\prime-t)
        \rho(t^\prime)_{m^\prime m_1^\prime}.
  \nonumber
\end{eqnarray}

\subsection{Energy representation}

Once the rules for calculating $\Sigma$ have been laid down in time domain, 
it is more convenient to shift into the energy representation with respect 
to the phonon energies and to do this in the $d$-basis.
The rule 2. remains as it is,
Eq.~(\ref{eq:correlator_E_app}) in the rule 3. is already in the energy 
representation, and we only need to reformulate the first and fourth rules.
\begin{enumerate}
\item
  {\bf Propagators}\\
  Assign a resolvent
  \begin{eqnarray}
    \frac{i}{\sigma\omega-E_{d^\prime}+E_d+i\eta}
  \end{eqnarray}     
  to each piece of a diagram temporally between two vertices, i.e., 
  irrespective of the propagator they lie on.
  Here $\omega$ is the phonon energy, and $E_{d^\prime}$ and $E_d$ are the
  energies of the states on the forward and backward propagators,
  respectively; $\sigma=+1$ $(-1)$ if the vertex with the earlier
  time lies on the forward (backward) propagator.
  The factor $\eta>0$ arises from the adiabatic turning on of the interaction
  and is taken to zero at the end.
\end{enumerate}
\begin{enumerate}
\setcounter{enumi}{3}
\item
  {\bf Summations and integrations}\\
  Sum over all internal indices and integrate over $\omega$.
  Due to the factor $i\eta$, the integrations are to be understood as 
  combinations of Cauchy's principal value integrals and delta functions.
\end{enumerate}

\begin{figure}
\epsfxsize=8.0cm
\centerline{\epsffile{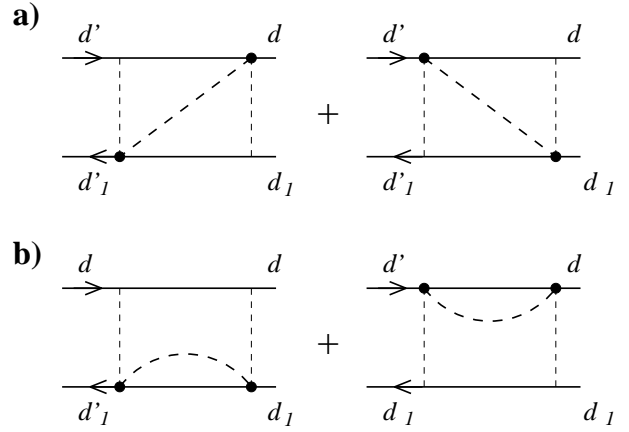}}
\caption{
  Diagrams may sometimes be pairwise combined to simplify the calculations. 
       }
\label{fg:diagram_pairs}
\end{figure}

In diagrams such as those in Fig.~\ref{fg:diagram_pairs}a 
the prefactors due to the vertices are equal. Therefore these can be 
pairwise combined and their integral parts written together as
\begin{eqnarray}
\label{eq:integrals_combined}
 && \int_{-\infty}^\infty d\omega\,
        \Gamma(\omega)\\
 && \;\;\;\;\;\;
        \cdot\left[\frac{i}{-\omega+(E_{d^\prime}-E_{d_1})+i\eta}
        + \frac{i}{\omega-(E_{d_1^\prime}-E_d)+i\eta}\right].\nonumber
\end{eqnarray}
If $d^\prime=d_1^\prime$ and $d=d_1$, i.e., if the density matrix is diagonal
before and after the transition, this expression simplifies into
\begin{eqnarray}
\label{eq:rate_diag}
 && \pi\int_{-\infty}^\infty d\omega\,
   \Gamma(\omega)
     \,\delta(\omega-(E_{d^\prime}-E_d)) \\ 
 &&
\;\;\;\;\;
   =\pi\, 
         \Gamma(E_{d^\prime}-E_d)
   =\frac{\pi A^2}{12\rho c^5\hbar^4}\cdot\frac{\Delta E^3}
                {e^{\beta\Delta E}-1}.
        \nonumber
\end{eqnarray}
In the last step we inserted the definition of $\Gamma(\omega)$ from
the equation (\ref{eq:Gamma}) and 
identified $\Delta E=E_{d^\prime}-E_d$ as the energy required for the
transition from the state $d$ to $d^\prime$. 
This expression, Eq.~(\ref{eq:rate_diag}), together with the prefactor
$G_{d^\prime d,d_1 d_1^\prime}$ is the phonon induced 
transition rate $\Sigma_{d,d^\prime}$ used in the literature and
in Parts~\ref{sec:dynamics} and \ref{sec:results} of this paper. 
From Eqs.~(\ref{eq:integrals_combined}) and (\ref{eq:rate_diag}) 
we see that the energy is only conserved 
(the delta function, cf. Fermi's golden rule) under some special 
circumstances and, in general, the rate need not be the simple real
function used in the literature.

\section{The spin-phonon rates}
\label{app:sigma}

In this Appendix, we outline the calculation of the phonon-induced transition 
rates $\Sigma$ between different spin states.
In order to calculate $\Sigma$ in lowest order in the spin-phonon coupling 
constants $g_i$, we need to evaluate the contractions
$\langle {\mathcal H}_{\rm sp}(t) {\mathcal H}_{\rm sp}(t^\prime)\rangle_{\rm ph}$,
where the expectation value is taken with respect to the phonon degrees of 
freedom.
Let us first insert Eq.~(\ref{eq:displacement}) into Eq.~(\ref{eq:Hsp}) 
and explicitly write down all the resulting terms
\begin{eqnarray}
  &&
    {\mathcal H}_{\rm sp}(t)=i \sum_{\vec{k}\sigma}
      \sqrt{ \frac{\hbar}{2MN\omega_{\vec{k}\sigma}} }
      [ b_{\vec{k}\sigma}^\dagger e^{-i\omega_{\vec{k}\sigma}t} 
           + b_{\vec{k}\sigma}^{} e^{i\omega_{\vec{k}\sigma}t}]
      e^{i\vec{k}\cdot\vec{r}}\nonumber\\
\label{eq:Hsp_app}
  &&\;\;\;\;\;\cdot    
\{g_1[S_x^2(t)-S_y^2(t)][e_x^{(\sigma)}k_x-e_y^{(\sigma)}k_y]\\
  &&\;\;\;\;\;\;\;\;
  +\frac{g_2}{4}\{S_x(t),S_y(t)\}[e_x^{(\sigma)}k_y+e_y^{(\sigma)}k_x]\nonumber\\
  &&\;\;\;\;\;\;\;\;
  +\frac{g_3}{4}(\{S_x(t),S_z(t)\}[e_x^{(\sigma)}k_z+e_z^{(\sigma)}k_x]\nonumber\\
  &&\;\;\;\;\;\;\;\;\;\;\;\;\;\;\;\;\;\;\;\;\;\;\;\;\;\;\;\;\;\;\;\;\;\;\;
   + \{S_y(t),S_z(t)\}[e_y^{(\sigma)}k_z+e_z^{(\sigma)}k_y])\nonumber\\
  &&\;\;\;\;\;\;\;\;
  +\frac{g_4}{4}(\{S_x(t),S_z(t)\}[e_x^{(\sigma)}k_z-e_z^{(\sigma)}k_x]\nonumber\\
  &&\;\;\;\;\;\;\;\;\;\;\;\;\;\;\;\;\;\;\;\;\;\;\;\;\;\;\;\;\;\;\;\;\;\;\;
   + \{S_y(t),S_z(t)\}[e_y^{(\sigma)}k_z-e_z^{(\sigma)}k_y]) \}.\nonumber
\end{eqnarray}
In evaluating the contraction, we need to consider all the possible states
before and after the action of ${\mathcal H}_{\rm sp}(t^{(\prime)})$ as well 
as all the possible orderings of the times $t$ and $t^\prime$ along the 
contour.
As a characteristic example, let us consider the diagram in 
Fig.~\ref{fg:rules}b and hold to the same usage of indices.
The time dependences of the spin operators are accounted for 
by the propagators and they can be put aside for the moment.
The remaining part with the spin operators is fully determined by the spin 
states before and after the vertices yielding
\begin{eqnarray}
  \langle\bar{m}_1|S^{\xi^\prime}|m_1\rangle
   \cdot\langle \bar{m}|S^\xi|m^\prime\rangle=
   \sqrt{S_{\bar{m},m^\prime} S_{m_1,\bar{m}_1}}.
\label{eq:spin_operators}
\end{eqnarray}
Here $\xi=\bar{m}-m^\prime,\xi^\prime=m_1-\bar{m}_1$ can take values
$\pm1$ or $\pm2$ due to the structure of ${\mathcal H}_{\rm sp}$, and, 
for brevity, $S^{\pm|\xi|}\equiv S_\pm^{|\xi|}$. 
The quantities on the right hand side of the equation are
\begin{eqnarray}
  S_{\bar{m},m^\prime}=(2m^\prime+\xi)
    \sqrt{S(S+1)-m^\prime(m^\prime+\xi)}
\end{eqnarray}
for $\xi=\pm1$ or
\begin{eqnarray}
  S_{\bar{m},m^\prime}&=&
     \left\{\left[S(S+1)-m^\prime(m^\prime+\nu)\right]\right.\\
     & &\;\;\;\;\;\;\;\;\;\;\left.
  \left[S(S+1)-(m^\prime+\nu)(m^\prime+2\nu)\right]\right\}^{1/2}
        \nonumber
\end{eqnarray}
for $\xi=\pm2$, $\nu=sign(\xi)$.

The part of the contraction depending on the phonon degrees of freedom
is evaluated assuming acoustic phonons with a linear dispersion relation 
and three modes indexed by $\sigma=1,2,3$ (one longitudinal and two 
transverse modes). The polarization vectors
$e_\alpha^{(\sigma)}$ are assumed unit vectors.
The resulting expression consists of two parts. The first one of them 
contains all the information concerning the phonon spectrum, energies, 
and temperature, 
\begin{eqnarray}
  \Gamma(t-t^\prime)&=&
  \int_{-\infty}^\infty d\omega\;\Gamma(\omega)\;e^{i\omega (t-t^\prime)}\\
  &\equiv&\int_{-\infty}^\infty d\omega 
          \frac{A^2}{12\rho c^5\hbar^4}\cdot
	\frac{\omega^3}{e^{\beta\omega}-1}\cdot e^{i\omega (t-t^\prime)}.
\label{eq:Gamma}
\end{eqnarray}
This part is defined such that it is independent of the spin states and
only contains the term $A^2$ from the coupling constants that turns out 
to be constant for all the rates.
In the diagrams, this corresponds to the spin-phonon interaction line. 
The second part, 
\begin{eqnarray}
\label{eq:Cs}
  C_{\xi,\xi^\prime}=
    \left\{\begin{array}{clrr}
                0, & {\rm for }|\xi|\neq|\xi^\prime|\\[4pt]
                1, & \xi=\xi^\prime=\pm1\\[4pt]
                \frac{15}{16}+\frac{1}{8}\delta_{\xi,\xi^\prime}, &
                        |\xi|=|\xi^\prime|=2
                \end{array}\right.,
\end{eqnarray}
however, does depend on the spin states and implies additional selection rules.
It should be noted that the third clause allows for $\xi\neq\xi^\prime$
if $|\xi|=|\xi^\prime|=2$.
For convenience, we combine the term $C_{\xi,\xi^\prime}$ with the spin 
operators and obtain the term 
\begin{eqnarray}
  G_{\bar{m} m^\prime,m_1 \bar{m}_1}=
        \sqrt{S_{\bar{m},m^\prime} S_{m_1,\bar{m}_1}}\cdot 
        C_{\xi,\xi^\prime},
\label{eq:Gs}
\end{eqnarray}
which in the diagrammatic language corresponds to a pair of vertices.

In the $d$-basis, the time dependence of the spin operators is of
a simple exponential form, cf. Eq.~(\ref{eq:propagator_md}), and 
the diagram in Fig.~\ref{fg:rules}b can be evaluated to yield
\begin{eqnarray}
\label{eq:example_rate_t}
 &&\Sigma(t-t^\prime)_{mm_1,m^\prime m_1^\prime}^{} = \\
 &&\;\;\;\;\sum_{\bar{m},\bar{m}_1}
    G_{\bar{m} m^\prime,m_1 \bar{m}_1}^{}
        \cdot\sum_{d,d_1}
        \langle m_1^\prime|d_1\rangle
        \langle d_1|m_1\rangle
        \langle m|d\rangle
        \langle d|\bar{m}\rangle\nonumber\\
 && \;\;\;\;\;\;\;\;\;\;\;\;\;\;\;\;\;\;\;\;\;\;\;\;\;\;\;\;\;\;\;\;\;\;\;\;\;
	\;\;\;\;\;\cdot e^{-i(E_d-E_{d^\prime})(t-t^\prime)}
        \;\Gamma(t^\prime-t).
  \nonumber
\end{eqnarray}

The actual transition rates are obtained by integrating 
Eq.~(\ref{eq:example_rate_t}) over the time difference $\tau=t-t^\prime$.  
This takes us to the energy representation, 
combining the exponential factor in Eq.~(\ref{eq:example_rate_t}) with the 
$e^{i\omega (t-t^\prime)}$ of the Fourier transform in Eq.~(\ref{eq:Gamma}).
The intergration over $\tau$ yields
\begin{eqnarray}
\label{eq:example_rate_w}
 &&\Sigma_{mm_1,m^\prime m_1^\prime}^{} = \\
 &&\;\;\;\;\sum_{\bar{m},\bar{m}_1}
    G_{\bar{m} m^\prime,m_1 \bar{m}_1}^{}
        \cdot\sum_{d,d_1}
        \langle m_1^\prime|d_1\rangle
        \langle d_1|m_1\rangle
        \langle m|d\rangle
        \langle d|\bar{m}\rangle\nonumber\\
 && \;\;\;\;\;\;\;\;\;\;\;\;\;\;\;\;\;\;\;\;\;\;\;\;\;\;\;\;\;\;\;\;\;\;\;\;\;
   	\cdot i\int_{-\infty}^\infty d\omega 
        \frac{\Gamma(\omega)}{-\omega-E_d+E_{d^\prime}+i\eta}.
  \nonumber
\end{eqnarray}

As the final step in calculating $\Sigma$, let us evaluate the integral in 
Eq.~(\ref{eq:example_rate_w}) or 
\begin{eqnarray}
  \tilde{I}_\sigma(\Delta E,\beta)&&=\int_{-\infty}^\infty d\omega
	\frac{\omega^3}{e^{\beta\omega}-1}
	\cdot\frac{1}{\sigma(\omega-\Delta E)+i\eta}\\[3pt]
  =&& P\int_{-\infty}^\infty d\omega
	\frac{\omega^3}{e^{\beta\omega}-1}
	\cdot\frac{1}{\sigma(\omega-\Delta E)}
  - i\pi\frac{\Delta E^3}{e^{\beta\Delta E}-1}\nonumber
\end{eqnarray}
for the general case. Here $\sigma=\pm1$.
The imaginary part of the integral (the last term) is, up to a prefactor, just 
$-i\pi\Gamma(\Delta E)$ and is independent of $\sigma$.
The real part in turn can be evaluated using the calculus of residues.
By combining the integral along the real axis with an infinite semicircle 
in the upper half plane to form a closed contour we obtain
${\rm Re}\{\tilde{I}_\sigma(\Delta E,\beta)\}=-2\pi{\rm Im}\sum\{enclosed\;{\rm residues}\}$.

The real part of the integral $\tilde{I}_\sigma(\Delta E,\beta)$ is 
divergent due to the $\omega^3$-term and some kind of a cutoff procedure 
is needed here. We have chosen a functional cutoff using instead of 
$\tilde{I}(\Delta E,\beta)$ the integral
\begin{eqnarray}
\label{eq:def_of_I}
  &&I_\sigma(\Delta E,\beta,D)=\\
  &&\;\;\;\;\;\;\int_{-\infty}^\infty d\omega
	\frac{\omega^3}{e^{\beta\omega}-1}
	\cdot\frac{1}{\sigma(\omega-\Delta E)+i\eta}
	\cdot\left(\frac{D^2}{D^2+\omega^2}\right)^2,\nonumber
\end{eqnarray}
i.e., the cutoff function is a Lorentzian squared and the additional parameter
$D$ is the cutoff parameter of the Lorentzian.
The integrand in (\ref{eq:def_of_I}) has single poles at 
$\omega=\Delta E-i\sigma\eta$, and at $\omega=2m\pi i/\beta$, 
where $m$ is a positive integer;
there are also second-order poles at $\omega=\pm iD$.
The residues $C^{(1)}$ from the first pole are real and can be neglected.
Collecting the other poles in the upper half plane and evaluating the residues 
yields
\begin{eqnarray}
  {\rm Re}\left\{I_\sigma(\Delta E,\beta,D)\right\}=
	-2\pi\left[\sum_{m=1}^\infty C_m^{(2)}+C^{(3)}\right],
\end{eqnarray}
where
\begin{eqnarray}
  &&\\[-5pt]
  &&C_m^{(2)}=\sigma\frac{\beta^2\Delta E D^4}{(2\pi)^3}
	\cdot\frac{m^3}{\left(\frac{\Delta E\beta}{2\pi}\right)^2+m^2}
	\cdot\frac{1}{\left[\left(\frac{D\beta}{2\pi}\right)^2-m^2\right]^2}
	\nonumber\\
  &&C^{(3)}=-\sigma\frac{D^5}{4(\Delta E^2+D^2)}
	\Bigg[
	\frac{\beta\Delta E}{4\sin^2\left(\frac{\beta D}{2}\right)}\\
  &&\;\;\;\;\;\;\;\;-\frac{1}{\Delta E^2+D^2}
	\left(\frac{3}{2}\Delta E^2+\frac{1}{2}D^2
		+\frac{\Delta E^3}{D}\cot\left(\frac{\beta D}{2}\right)\right)
	\Bigg].\nonumber
\end{eqnarray}
If $D$ is accidentally chosen to equal $2\pi m^*/\beta$ for some $m^*$,
the residues $C_{m^*}^{(2)}+C^{(3)}$ should be replaced by
\begin{eqnarray}
  C^*=\sigma\frac{D^3}{16\beta}
	\cdot\frac{3\Delta E^5-14\Delta E^3D^2-\Delta ED^4}
		  {(\Delta E^2+D^2)^3}.
\end{eqnarray}

In the $d$-basis and using only the diagonal elements of the reduced density 
matrix, the integrals $I_\sigma(\Delta E,\beta,D)$ can always be combined such
that their real parts cancel each other.
When using the nondiagonal elements of $\rho(t)_{d,d^\prime}$, as well,
this is no longer true and the real parts of $I_\sigma(\Delta E,\beta,D)$
give rise to $D$-dependent shifts in the energies $E_d$.
In this work, we have used $D$'s of the order of the anisotropy barrier, 
i.e., 50-100K, and found that this yields tiny shifts also in the relaxation 
rate curves but leaves the qualitative picture unchanged.
In the figures in the section \ref{sec:results},
$D$ is chosen as low as 25K in order to simplify the comparison between
the calculations with and without the nondiagonal elements.

With all the contributions to $\Sigma$ written down, we can find an 
estimate for the order of magnitude of the elements (the individual rates)
$\Sigma_{mm_1,m^\prime m_1^\prime}$.
The most interesting piece of information for each state is the {\it largest 
rate} coupling that state to other states -- this rate plays a key role in
justifying the neglect of the nondiagonal states in $\rho(t)$,
see App.~\ref{app:basis}, as well as in the suppression of the narrow 
resonances found in the text.

The prefactor in $\Gamma(\omega)$, cf. Eq.~(\ref{eq:Gamma}),
amounts to $7.0\cdot 10^5 c^{-5}{\rm s}^5{\rm m}^{-5}{\rm K}^{-2}$,
where the sound velocity $c$ is expressed in meters per second.
The units are chosen such that, when $\omega$ is expressed in kelvin, 
also $\Gamma(\omega)$ is given in kelvin.
For $\omega>0$ (and also $\omega>k_{\rm B}T$), i.e., for transitions 
related with phonon absorption, the energy-dependent part of $\Gamma(\omega)$ 
strongly decreases for increasing $\omega$; 
for $\omega<0$, corresponding to phonon emission, $\Gamma(\omega)$ approaches 
the temperature-independent power-law dependence $\omega^3$.
The largest $\Gamma(\omega)$'s are attained for these latter processes
in connection with low-energy spin states. 
The contribution from the spin operators, cf. Eq.(\ref{eq:spin_operators}),
on the other hand, is larger for spin states closest to the top
of the barrier and tends to balance the changes in $\Gamma(\omega)$
and reduce the variations in $\Sigma_{mm_1,m^\prime m_1^\prime}$ for 
different states and for varying $H_z$.
The typical energy scale arising from the spin-phonon rates
is found to be $10^{-5}-10^{-4}$K.

\section{Choice of basis}
\label{app:basis}

In sections \ref{sec:diagrams} and \ref{sec:kinetic}, we decidedly 
formulated the more general equations independent of the chosen basis for 
${\mathcal H}_S$. In this appendix, we consider the eigenbases of 
${\mathcal H}_S$ or the $d$-basis, in more detail.

The (strong) tunneling poses a problem for the diagrammatic formulation
in the $m$-basis,
but this can be easily solved by first diagonalizing ${\mathcal H}_S$ and 
then expressing all the equations in its eigenbasis.
In this $d$-basis, the kinetic equation for the diagonal and off-diagonal 
density matrix elements reads
\begin{eqnarray}
\label{eq:kinetic_d_diag}
  \dot{\rho}(t)_{d,d}&=&
        \sum_{d_1,d_1^\prime} \!\Sigma_{dd,d_1 d_1^\prime}
        \rho(t)_{d_1,d_1^\prime}
\end{eqnarray}
and
\begin{eqnarray}
\label{eq:kinetic_d_offdiag}
  \dot{\rho}(t)_{d,d^\prime}&=&-i(E_d\!-\!E_{d^\prime})\rho(t)_{d,d^\prime}
        \!+\!\!\sum_{d_1,d_1^\prime} \!\Sigma_{d d^\prime,d_1 d_1^\prime}
        \rho(t)_{d_1,d_1^\prime},
\end{eqnarray}
respectively.
From the knowledge of (the full) $\rho(t)$ we can again obtain, e.g., 
the magnetization
\begin{eqnarray}
  M(t)&=&g\,\mu_{\rm B}\sum_{d,d^\prime}\sum_m 
 \,\langle d^\prime|m\rangle \,m\, \langle m|\,d\rangle \,\rho_{d,d^\prime}(t)
\\
 &\equiv&g\,\mu_{\rm B}\sum_{d,d^\prime} m_{d,d^\prime} \rho_{d,d^\prime}(t)
\end{eqnarray}
where $m_{d,d^\prime}$ is defined in this way as the matrix element of $S_z$
in the $d$-basis.

\begin{figure}
\epsfxsize=8.0cm
\centerline{\epsffile{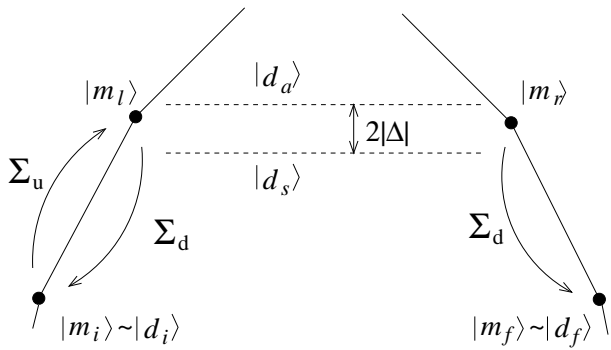}}
\caption{
  Schematic of the situation discussed in the text. 
  If the spin-phonon coupling denoted by the arrows and $\Sigma$'s 
  is much weaker than the tunnel coupling $\Delta$ between the states 
  $|m_l\rangle$ and $|m_r\rangle$, these states can be thought of as 
  an effective two-state system. The eigenstates of the two-state system 
  are depicted as dashed lines separated in energy by $2|\Delta|$.
  The subscripts of the $\Sigma$'s correspond to those used in the text.
       }
\label{fg:figto_app_basis}
\end{figure}

When the tunneling rates dominate the kinetic equations, the main features of 
the eigenstates can be understood in even simpler terms as follows.
When a given state is off-resonant, there is essentially a one-to-one
correspondence between each of the $m$- and $d$-states, i.e., also
the $d$-states are localized on one or the other side of the barrier.
Close to a resonance, two states $m_l$ and $m_r$ on different sides 
of the barrier, see Fig.~\ref{fg:figto_app_basis},
get coupled and form an approximate two-state system described by 
\begin{eqnarray}
  {\mathcal H}_2=
        \left(\begin{array}{clrr}
          E_{m_l} & \Delta \\
          \Delta^* & E_{m_r}
        \end{array}\right).     
\label{eq:two-state-system}
\end{eqnarray}
The nondiagonal elements denote the tunnel splitting as obtained from 
the diagonalization of the full spin Hamiltonian;
the subscripts stand for the left and right sides of the barrier.
The eigensolutions to this are the symmetric and antisymmetric combinations 
of the respective $m$-states
\begin{eqnarray}
\label{eq:alpha_beta1}
  |d_s\rangle &=& \alpha|m_l\rangle+\beta|m_r\rangle\\
  |d_a\rangle &=& \beta|m_l\rangle-\alpha|m_r\rangle
\label{eq:alpha_beta2}
\end{eqnarray} 
that extend through the barrier, see Fig.~\ref{fg:figto_app_basis}.
The factors $\alpha$ and $\beta$ are the normalized constants
\begin{eqnarray}
  \alpha&=&\frac{\Delta}{\sqrt{\hat{\varepsilon}^2+|\Delta|^2}}\\
  \beta&=&\frac{\hat{\varepsilon}}{\sqrt{\hat{\varepsilon}^2+|\Delta|^2}}
\end{eqnarray}
with $\hat{\varepsilon}=\frac{1}{2}[(E_l-E_r)-\sqrt{(E_l-E_r)^2+4|\Delta|^2}]$.

The biggest simplification is attained when we argue that, for the most 
values of $H_z$, we can restrict our considerations to the diagonal 
elements of the density matrix. 
A naive justification for this concerns the stationary values of the density 
matrix elements [obtained by requiring $\dot{\rho}(t)_{d,d^\prime}=0$]. 
This leads to the immediate 
conclusion that all the off-diagonal elements between nonresonant
states are negligibly small. Furthermore, the nondiagonal elements
are also very small for any pair of resonant states as long as the tunnel 
splitting of that particular resonance is larger than the spin-phonon rates 
coupling these states to others, see the end of App.~\ref{app:sigma}.

We also investigated the temporal behaviour of the off-diagonal elements
in terms of the reduced model shown in Fig.~\ref{fg:figto_app_basis} and
the results lend support to the above conclusions.
The idea of this simulation was to prepare the system into the state
$d_i$ at the initial time $t_0$, let the system then evolve in time
according to the kinetic equation, and see how the off-diagonal elements
$\rho(t)_{d_s,d_a}$ and $\rho(t)_{d_a,d_s}$ behave.
The resonant pair of states in the figure is similar to
the one in Eqs.~(\ref{eq:alpha_beta1}) and (\ref{eq:alpha_beta2})
and it is coupled to two lower, nonresonant states $d_i$ and $d_f$. 
The rates depicted in the figure are $\Sigma_u=\Sigma_{ll,ii}$ and
$\Sigma_d\approx(\Sigma_{ii,ll}+\Sigma_{ff,rr})/2$.
The magnitudes of these rates -- as compared to the tunnel splitting 
$|2\Delta|$ -- determine two regimes.
If $2|\Delta|\gg\Sigma_{\rm d}$, the amplitudes of the nondiagonal elements 
are found to quickly reach their maxima $\propto\Sigma_{\rm u}/|2\Delta|$ and 
their values orbit around and ``decay'' towards the respective complex 
stationary values. 
On the other hand, according to the detailed-balance relation, 
the stationary values of the diagonal elements 
are proportional to $\Sigma_{\rm u}/\Sigma_{\rm d}$. Hence
$\rho(t)_{d,d^\prime}/\rho(t)_{d^{(\prime)}}\propto\Sigma_{\rm d}/|2\Delta|$
and we can neglect the nondiagonal elements, if $\Sigma_{\rm d}\ll 2|\Delta|$.
In this case the kinetic equation, Eqs.~(\ref{eq:kinetic_d_diag}) and 
(\ref{eq:kinetic_d_offdiag}), becomes very simple: 
Eq.~(\ref{eq:kinetic_d_offdiag}) can be neglected and the rate 
$\Sigma$ acquires the form, cf. App.~\ref{app:sigma}, 
\begin{eqnarray}
  \Sigma_{d^\prime d^\prime,dd}^{}=
    \pm G_{d d^\prime,d^\prime d}\,
        \frac{\pi A^2}{12\rho c^5\hbar^4}\cdot\frac{\Delta E^3}
                {e^{\beta\Delta E}-1},
\end{eqnarray}
where
\begin{eqnarray}
  G_{d d^\prime,d^\prime d}&=&\!\sum_{m_1,m_2}\sum_{m_3,m_4} 
        \!\!\!G_{m_4m_3,m_2m_1}\\[-4pt]
&&\;\;\;\;\;\;\;\;\;\;\;\;\;\;\;\;\;\;\;\;\;\;\;\;\;
        \cdot\langle d^\prime| m_4\rangle \langle m_3| d\rangle 
        \langle d|m_2\rangle \langle m_1|d^\prime \rangle\nonumber
\end{eqnarray}
corresponds to the vertices and contains the spin-phonon 
coupling constants, see App.~\ref{app:rules}.
In the opposite case, $2|\Delta|\ll\Sigma_{\rm d}$, the nondiagonal elements 
do not perform orbiting motion in the complex plane but increase 
motonously to roughly one half of $\rho_{d^{(\prime)}}^{(0)}$.
In this case, the off-diagonal elements clearly cannot be neglected.

For Mn$_{12}$ we can attain the whole range of cases: for the most strongly 
coupled level(s) $2|\Delta|\gg\Sigma_{\rm d}$, while for the lower levels, 
$2|\Delta|\ll\Sigma_{\rm d}$. In the former case, the diagonal elements 
$\rho_d(t)$ are sufficient in describing the system whereas, 
in the latter case, we either have to include the nondiagonal states or 
restrict our considerations to magnetic fields for which 
$\Delta E\gg\Sigma_{\rm d}$ for all the levels, 
cf. Ref.~\onlinecite{Luisetal2}.
In the text, we neglect the nondiagonal elements in the calculation 
of $\chi(\omega)$ and in some of the analytical considerations 
but compare the two cases in section \ref{sec:results}.

\section{Laplace transformation}
\label{app:Laplace}

In this appendix, we consider the Laplace tranformation 
\begin{eqnarray}
  f(z)\equiv\int_0^\infty \!\!dt\, e^{-izt} f(t)
\end{eqnarray}
of the kinetic equation, Eq.~(\ref{eq:kinetic}).
We also give another proof of the applicability of the Markov 
approximation in calculating the relaxation rates.

The kinetic equation is readily transformed into
\begin{eqnarray}
  -iz\rho(z)-\rho(t&=&0)=-iL_0\,\rho(z)+\Sigma(z)\rho(z)\\[5pt]
 \Rightarrow \rho(z) &=& \frac{\rho(t=0)}{-iz+iL_0-\Sigma(z)}.
\label{eq:poles}
\end{eqnarray}
The poles of Eq.~(\ref{eq:poles}), i.e., the solutions of 
$-iz_i+iL_0-\Sigma(z_i)=0$, yield the exact eigenvalues to the kinetic 
equation: $z_i=\omega_i+i/\tau_i$. For the slowest mode of the time 
evolution, one can consider the expansion
\begin{eqnarray}
  \Sigma(z_1)\approx\Sigma(0)
        +z_1\cdot\frac{\partial \Sigma(z)}{\partial z}\Big\vert_{z=0}
        +...
\end{eqnarray}
The prefactor of the $z_1$ may be evaluated to be proportional
to $[\tau_1\cdot\min\{\Delta E,k_{\rm B}T,D\}]^{-1}$, i.e., to the maximal
ratio between the over-barrier relaxation rate $1/\tau_1$ and the other 
characteristic energy scales in the problem: level spacing and/or splitting 
$\Delta E$, temperature $k_{\rm B}T$, and the cutoff of the phonon spectrum 
$D$ (the cutoff $D$ is introduced in order to assure that the real part 
of Eq.~(\ref{eq:integrals_combined}) is convergent also when we consider
the nondiagonal density matrix elements). 
It turns out that the actual relaxation rates are several orders of magnitude 
smaller than any other energy scale and it becomes safe to approximate
\begin{eqnarray}
  \rho(z)\approx \frac{\rho(t=0)}{-iz+iL_0-\Sigma(0)}=\frac{\rho(t=0)}{-iz-W}
\label{eq:Laplace_Markov}
\end{eqnarray}
where $\Sigma(0)$ has been identified as the constant $\Sigma$ of 
the Markov approximation above and $W$ is defined accordingly,
cf. Eq.~(\ref{eq:kinetic_W}).
This approximation is valid for the relaxation mode and time $\tau_1$
\begin{eqnarray}
  \rho^{(1)}(z)=\frac{\rho^{(1)}(t=0)}{-iz-1/\tau_1}
\end{eqnarray}
but the Markov approximation may give erroneous results for 
the faster eigenmodes for which Eq.~(\ref{eq:Laplace_Markov}) no longer 
holds true.

\section{Lorentzian peak shapes}
\label{app:Lorentz}

The series of peaks found in the relaxation rates/times, cf. 
Fig.~\ref{fg:relaxation}, may be understood in terms of different relaxation
paths, each path with a possible tunneling channel giving rise to a peak --
see \onlinecite{LL} for nice illustrations of the paths.
In this appendix, we sketch a derivation that aims to show that 
the Lorentzian peak shapes are actually something to be expected.

When the spin system is somehow disturbed away from equilibrium
and then let relax, it quickly acquires a metastable state, 
a thermal equilibrium separately on each side of the barrier.
This initial thermalization into the metastable state is driven by 
the spin-phonon interaction that can change the spin states $m$ by
$\pm1$ or $\pm2$.
At a much longer time scale, the system relaxes over the barrier
towards the real equilibrium configuration.
For the relaxation to take place, the crucial step is the final
transition that transfers the spin onto the other side of the barrier. 
We can distinguish two regimes in terms of how this critical transition 
takes place.
In absence of tunneling, e.g., in off-resonance conditions, 
the relaxation is only possible over the top of the barrier, while
for relatively strong tunnel splitting and for resonant conditions, 
the dominant path is via tunneling across the barrier well below its top.
When the tunneling is weak compared with the spin-phonon interaction, 
the tunneling rate is the bottle neck for the relaxation to take place.
In Mn$_{12}$, this is the case for tunneling between the low-lying states
with $|m|>4$ (for $H_z\approx0$T).
However, for the experimentally relevant resonances, the tunneling is 
strong and takes place between the higher states.
In this case, the spin actually oscillates back and forth through 
the barrier until it relaxes to some lower state on 
either side of the barrier. This is the case of interest here.

\begin{figure}
\epsfxsize=8.5cm
\centerline{\epsffile{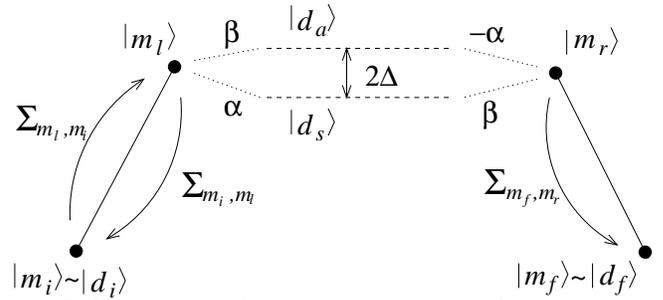}}
\caption{
  Illustration of the parameters discussed in the text:
  states in the $m$ and $d$-bases, transition rates $\Sigma_{m,m^\prime}$,
  and factors $\pm\alpha$ and $\beta$ from Eq.~(\ref{eq:two-state-system}).
       }
\label{fg:figto_app_Lorentz}
\end{figure}

In the strong-tunneling regime, the system is best described in terms of 
the $d$-basis where it suffices to consider the diagonal elements of
the density matrix.
In order to get a more intuitive picture of the relaxation, let us 
consider a situation where the system has been prepared onto one side 
of the barrier and has reached the metastable thermal equilibrium there.
This initial condition is convenient for two purposes: first, the relaxation
only proceeds into one direction and, second, the phonon-induced transitions 
on this one side of the barrier are accounted for by the thermal 
probabilities $\tilde{\rho}_{d}$ (tilde denotes the metastable state and 
we write just one index for the diagonal matrix elements).
Let us further consider relaxation via a single tunneling resonance
and take into account the states and transition processes illustrated
in Fig.~\ref{fg:figto_app_Lorentz}.
The system starts in the initial state $d_i$ (localized onto the left side of
the barrier, $\sim m_i$), is then activated onto resonance 
into either the symmetric or antisymmetric state, denoted by $d_s$ and $d_a$, 
respectively, and at some point is transfered down to the final state $d_f$ 
(localized onto the right side of the barrier, $\sim m_f$). 
The subsequent intravalley relaxation is so fast that after the transition 
to $d_f$ the relaxation can be considered complete.
The states $d_s$ and $d_a$ extend through the barrier and this is the key 
point of the present discussion: the spin is transferred through the barrier 
in a single step, the rate being determined by thermal activation
but also by the magnetic-field dependent amplitudes $\alpha$ and $\beta$, 
from Eqs.~(\ref{eq:alpha_beta1}) and (\ref{eq:alpha_beta2}), for the two 
extended states to be on either side of the barrier.
These amplitudes determine the relative probabilities for the activation
process to couple to the resonant states.

The above discussion can be formulated in the language of a master equation:
\begin{eqnarray}
  \dot{\tilde{\rho}}_{d_s}(t)=0&\approx& 
        \Sigma_{d_s,d_i}\tilde{\rho}_{d_i}(t) - (\Sigma_{d_f,d_s}
        +\Sigma_{d_i,d_s})\tilde{\rho}_{d_s}(t)\\
        &&\!\!\!\!\!\!\!\!\!\!
       \approx |\langle d_s|m_l\rangle|^2 \Sigma_{m_l,m_i}\tilde{\rho}_{d_i}(t)
        \nonumber\\
        &&\!\!\!\!\!\!\!\!\!\!
        -(|\langle d_s|m_r\rangle|^2 \Sigma_{m_f,m_r}
          \!+ |\langle d_s|m_l\rangle|^2 \Sigma_{m_i,m_l})\tilde{\rho}_{d_s}(t)
        \nonumber\\
        &&\!\!\!\!\!\!\!\!\!\!= 
        |\alpha|^2 \Sigma_{m_l,m_i}\tilde{\rho}_{d_i}(t)\nonumber\\[2pt]
        &&\;\;\;\;\;\;\;\;\;\;-(|\beta|^2 \Sigma_{m_f,m_r}
          + |\alpha|^2 \Sigma_{m_i,m_l})\tilde{\rho}_{d_s}(t)\nonumber\\[8pt]
  \dot{\tilde{\rho}}_{d_a}(t)=0&\approx& \Sigma_{d_a,d_i}\tilde{\rho}_{d_i}(t)
        - (\Sigma_{d_f,d_a} + \Sigma_{d_i,d_a})\tilde{\rho}_{d_a}(t)\\
        &&\!\!\!\!\!\!\!\!\!\!
       \approx |\langle d_a|m_l\rangle|^2 \Sigma_{m_l,m_i}\tilde{\rho}_{d_i}(t)
        \nonumber\\
        &&\!\!\!\!\!\!\!\!\!\!
        -(|\langle d_a|m_r\rangle|^2 \Sigma_{m_f,m_r}
         \!+ |\langle d_a|m_l\rangle|^2 \Sigma_{m_i,m_l})\tilde{\rho}_{d_a}(t)
        \nonumber\\
        &&\!\!\!\!\!\!\!\!\!\!
        = |\beta|^2 \Sigma_{m_l,m_i}\tilde{\rho}_{d_i}(t)\nonumber\\[2pt]
        &&\;\;\;\;\;\;\;\;\;\;-(|\alpha|^2 \Sigma_{m_f,m_r}
              +|\beta|^2 \Sigma_{m_i,m_l})\tilde{\rho}_{d_a}(t).\nonumber
\end{eqnarray}
The approximate equalities are just a reminder that we have neglected, e.g., 
the contributions from states above the resonance as well as the return 
possibility from state $d_f$.
These equations can be simplified by the assumption 
$\Sigma_{m_i,m_l}\approx\Sigma_{m_f,m_r}$ which is reasonable for
a pair of resonant states -- as a result, the $\Sigma$'s can be taken out of
the brackets in the last forms of the above formulas. By further noting
that due to normalization $|\alpha|^2+|\beta|^2=1$ and that
the resulting probabilities are time independent, we obtain
\begin{eqnarray}
  \tilde{\rho}_{d_s}&\approx&
      |\alpha|^2\,\frac{\Sigma_{m_l,m_i}}{\Sigma_{m_i,m_l}}\,\tilde{\rho}_{d_i}
        \nonumber\\
  \tilde{\rho}_{d_a}&\approx&
      |\beta|^2\,\frac{\Sigma_{m_l,m_i}}{\Sigma_{m_i,m_l}}\,\tilde{\rho}_{d_i}.
        \nonumber
\end{eqnarray}
The ratio of the $\Sigma$'s is just the thermal factor 
$\exp[-\beta(E_l-E_i)]$, cf. detailed balance, and $\tilde{\rho}_{d_i}$ is
the thermal probability to be in a state with energy $E_i$ over the lowest 
energy $E_{-10}$ on left hand side (for $H_z>0$).
Together these yield a factor $c\cdot\exp[-\beta(E_l-E_{-10})]$,
where $c$ is a normalization constant equal to $\tilde{\rho}_{-10}$
which is close to unity for the temperatures of interest.

In the next and final step, the relaxation rate is obtained
from the knowledge of these probabilities and the rates to be dragged
down on the right hand side of the barrier, i.e., 
\begin{eqnarray}
  \tau^{-1}&\approx& \Sigma_{d_f,d_s}\cdot\tilde{\rho}_{d_s}
                +\Sigma_{d_f,d_a}\cdot\tilde{\rho}_{d_a}\nonumber\\[2pt]
        &\approx& |\beta|^2 \Sigma_{m_f,m_r}\,\tilde{\rho}_{d_s}
               +|\alpha|^2 \Sigma_{m_f,m_r}\,\tilde{\rho}_{d_a}
                \nonumber\\[2pt]
        &\approx& 2\,|\alpha|^2|\beta|^2 
                \,\Sigma_{m_f,m_r}\,c\cdot e^{-\beta(E_i-E_{-10})}.
\label{eq:rate_simple}  
\end{eqnarray}
The exponential factor is just the effective Arrhenius factor seen in 
experiments, $\frac{1}{2}\,c\,\Sigma_{m_f,m_r}=\tau_0^{-1}$, and 
\begin{eqnarray}
  4\,|\alpha|^2|\beta|^2 = \frac{(2|\Delta|)^2}
        {(E_{m_l}-E_{m_r})^2+(2|\Delta|)^2}.
\label{eq:Lorentz}
\end{eqnarray}
Here $2|\Delta|$ is the tunnel splitting.
It is more or less independent of $H_z$ but 
$\xi\equiv E_{m_l}-E_{m_r}$ can be tuned with the magnetic field. 
In terms of the field, the width of the resonant peak at its half maximum is
\begin{eqnarray}
  \delta H_z=\frac{4|\Delta|}{g\mu_{\rm B}|m_l-m_r|}.
\end{eqnarray}

This sketch of a derivation introduces all the factors seen in 
experiments: the Arrhenius law with a reasonable prefactor $\tau_0^{-1}$,
that depends weakly on temperature and the particular resonance,
see the two last paragraphs of App.~\ref{app:sigma},
and peaks of accelerated relaxation superimposed on it.
The peak heights or the relaxation rates on resonance are found to
correspond to the Boltzmann or Arrhenius factor with the energy 
corresponding to 
the effective barrier height. The peak shape is Lorentzian as observed 
in experiment with widths given by precisely the tunnel splittings.

\pagebreak


\begin{references}
\bibitem{is_this_so}
The division between thermally-activated and tunneling-enhanced relaxation 
depends on typical experimental time scales.
In relaxation measurements, the magnetization first relaxes 
faster than exponentially and only asymptotically reaches 
an exponential regime. 
For $T>3{\rm K}$, almost all the spins relax before reaching this stage.
In susceptibility experiments, on the other hand, the relevant time scale
is the measurement frequency and this can keep up with the relaxation rate
up to 5-6K.
\bibitem{fe8}
This bears analogy to the related molecular magnet Fe$_8$, which has 
hard, medium, and easy axes -- 
the interference effects in the tunnel splittings appear when the external
magnetic field is oriented parallel to the hard axis.
\bibitem{berries}
Similar results were recently reported in Ref.~\onlinecite{review};
see Ref.~\onlinecite{LL_Berry} for a detailed analysis of 
the interference effects.
\bibitem{comment_on_Alsaqer}
In Ref.~\onlinecite{Alsaqeretal} the authors calculate the energy
spectrum for Mn$_{12}$ starting from the twelve atomary spins and find
different level splittings to those predicted by the simple $S=10$ model. 
The discrepancy varies between different pairs of states and 
as ratio varies between 0.8 to 5 from higer to lower states, respectively.
Such details do not affect the essence of this paper and may be inserted
as parameters into the calculation.
\bibitem{Markov}
Rigorously taken, instead of $\rho(t^\prime)\approx\rho(t)$, we should use 
$\rho(t^\prime)=e^{i{\mathcal H}_S(t-t^\prime)}\rho(t)e^{-i{\mathcal H}_S(t-t^\prime)}$ 
and insert this into Eq.~(\ref{eq:kinetic}). 
Combining the exponential factors yields the expressions of 
Refs.~\onlinecite{Blum_book} and \onlinecite{LL}. 
This procedure complicates the correspondence of the indices in the equations
and in the diagrams, and, as this is found to make very little difference 
to the final results,
\cite{we_did_this}
we neglect this step here.
\bibitem{we_did_this}
We performed the calculation with both kinds of Markov approximations
and the results did not show any significant differences.
It is also noteworthy that when using only the diagonal states of 
the eigenbasis the two approximations become equal.
\bibitem{full_matrix}
We have solved the problem numerically for the full density matrix in 
both bases, i.e., either first diagonalizing the spin part or taking 
the tunneling matrix elements into account to all orders in the $m$-basis.
The results are of course equivalent.
\bibitem{decoherence}
The discussion on the decoherence is relevant for the low-lying states
with small tunnel splittings.
\end{references}
\end{document}